\documentclass[preprint2]{aastex}
\usepackage{amsmath}
\usepackage{multirow}
\pdfpageattr{/Group <</S /Transparency /I true /CS /DeviceRGB>>}
\graphicspath{{figures/}}
\defcitealias{Boue_Laskar_Icarus_2006}{BL06}
\defcitealias{Boue_Laskar_Icarus_2009}{BL09}
\defcitealias{Kaib_etal_ApJ_2011}{K11}
\newcommand{\mtwo}[1]{\multicolumn{2}{l}{#1}}
\renewcommand{\vec}[1]{{\boldsymbol #1}}
\newcommand{\lap}[2]{{\rm b}_{#1}^{(#2)}}

\newcommand{\ii}{{\rm i}}

\def\abs#1{\left\vert#1\right\vert}
\def\norm#1{\left\Vert#1\right\Vert}

\newcommand{\grad}[1]{\boldsymbol{\nabla}_{\!#1}}
\newcommand{\moy}[2]{\left\langle{#2}\right\rangle_{#1}}
\def\crm{\cr\noalign{\medskip}}

\def\m@th{\mathsurround=0pt}
\def\EQM#1{\vcenter{\normalbaselines\m@th
    \ialign{${\displaystyle ##}$\hfil&&\ ${\displaystyle ##}$\hfil\crcr
    \mathstrut\crcr\noalign{\kern-\baselineskip}
    \noalign{\smallskip}
    #1\crcr\mathstrut\crcr\noalign{\kern-\baselineskip}}}}

\newcommand{\be}{\begin{equation}}
\newcommand{\ee}{\end{equation}}

\def\Dron#1#2{\frac{\partial#1}{\partial#2}}

\newcommand{\bpm}{\begin{pmatrix}}
\newcommand{\epm}{\end{pmatrix}}
\newcommand{\figKaiba}{
\begin{figure}[t]
\plotone{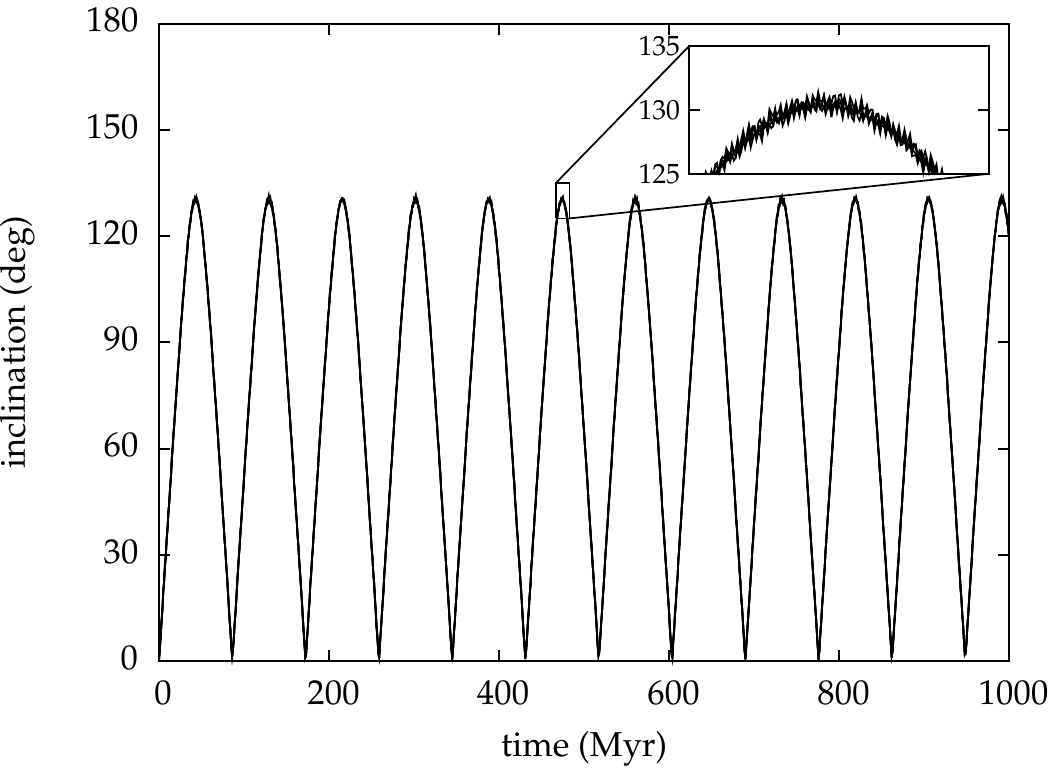}
\caption{\label{fig.Kaiba}Evolution of the inclination of all the planets of
the 55 Cancri system vs. time obtained with the numerical model.
Inclination is measured with respect to the initial planetary orbital plane.
The inset plot resolves the evolution of each planet's inclination.}
\end{figure}
}
\newcommand{\figKaibb}{
\begin{figure*}
\includegraphics[width=\linewidth]{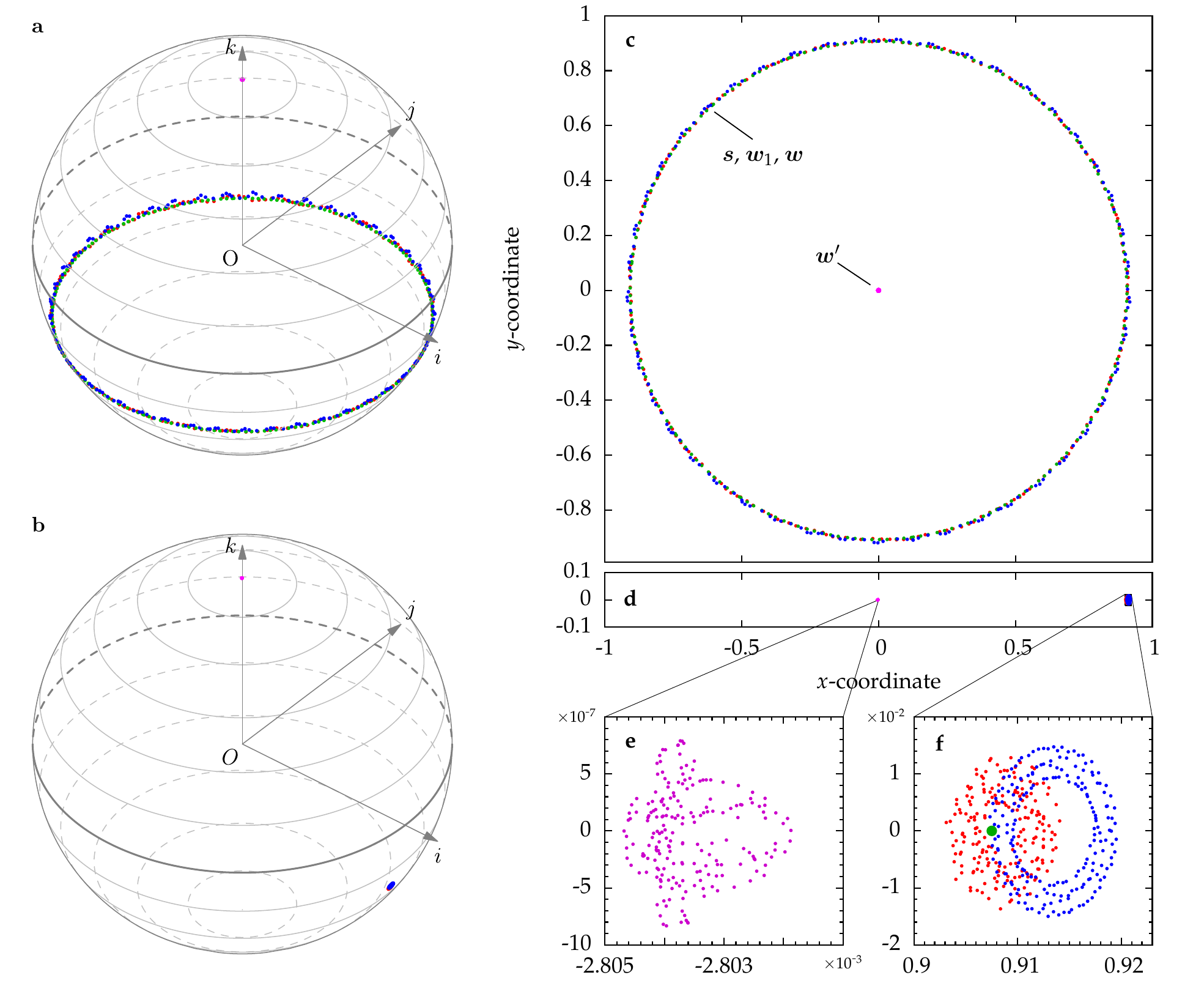}
\caption{\label{fig.Kaibb}Trajectory of the stellar spin axis $\vec s$
(blue), and of the unit orbital angular momentum of the
innermost planet $\vec w_1$ (red), of all the planets $\vec w$
(green), and of the stellar companion $\vec w'$ (magenta) in the
55~Cancri system. {\bf (a)} Representation in a fixed reference frame on
the unit sphere whose north pole coincides with the direction of the
total angular momentum of the system. {\bf (b)} Representation in a
frame rotating with the main precession frequency.  {\bf (c)} Projection
of the trajectory on the $x$-$y$ plane in the fixed reference frame.
{\bf (d)} Projection on the $x$-$y$ plane in the rotating frame.  {\bf
(e)} Zoom on the trajectory of $\vec w'$ in the $x$-$y$ plane. {\bf (f)}
Zoom on the trajectory of $\vec s$, $\vec w_1$, and $\vec w$ (large
green point) in the $x$-$y$ plane.  The initial conditions are the same
as in Fig.~\ref{fig.Kaiba}.}
\end{figure*}
}
\newcommand{\figKaibc}{
\begin{figure*}[t]
\includegraphics[width=\linewidth]{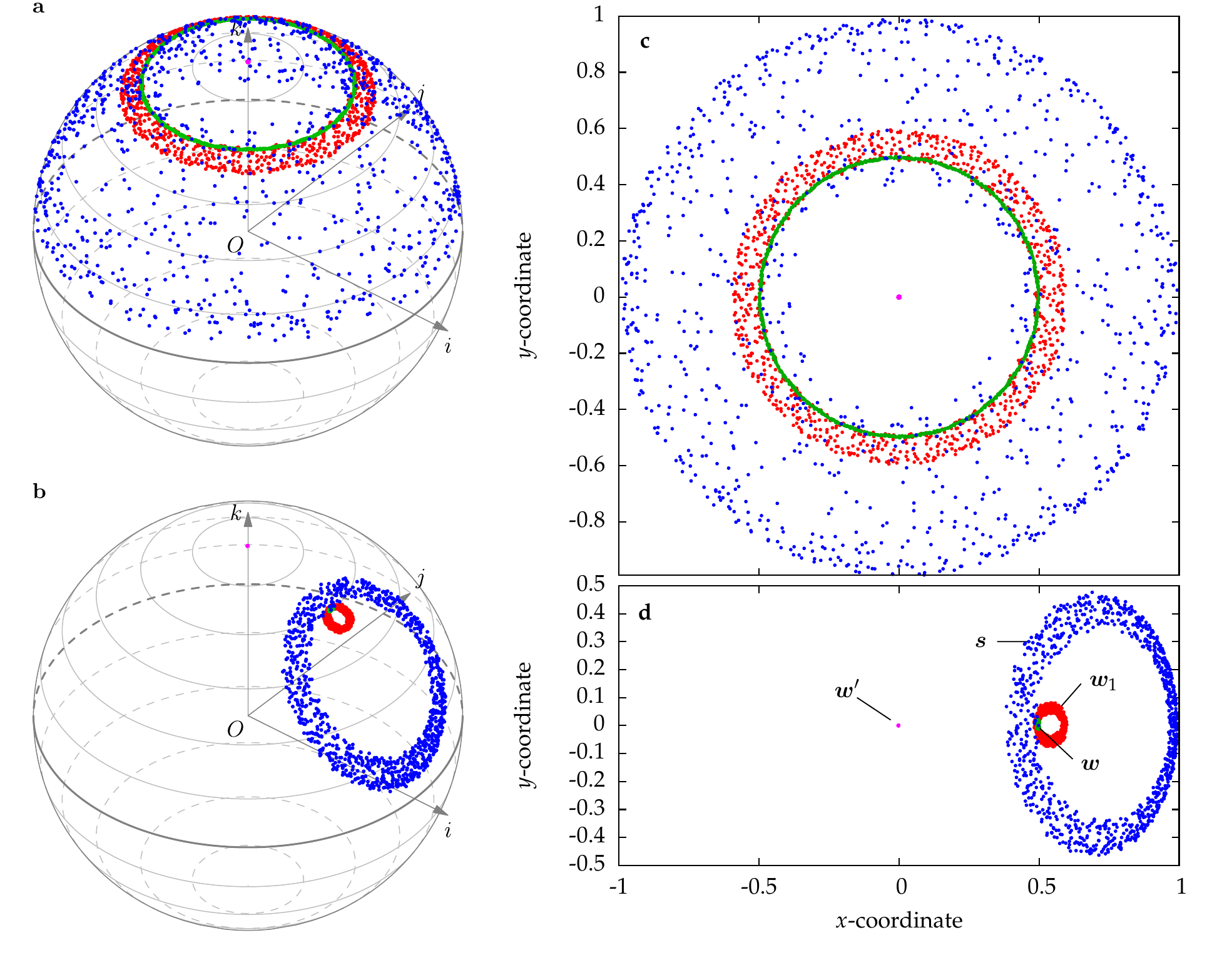}
\caption{\label{fig.Kaibc}Same as Fig.~\ref{fig.Kaibb} but with the
stellar companion 55~Cnc~B on a closer and less inclined orbit: $a'=550$
au, $e'=0.936$, and $I'=30^\circ$, initially. In panel {\bf d}, $\vec
w_2$, $\vec w_3$, and $\vec w_4$ follow $\vec w_1$, while $\vec w_5\sim
\vec w$.}
\end{figure*}
}
\newcommand{\figHDa}{
\begin{figure}[t]
\plotone{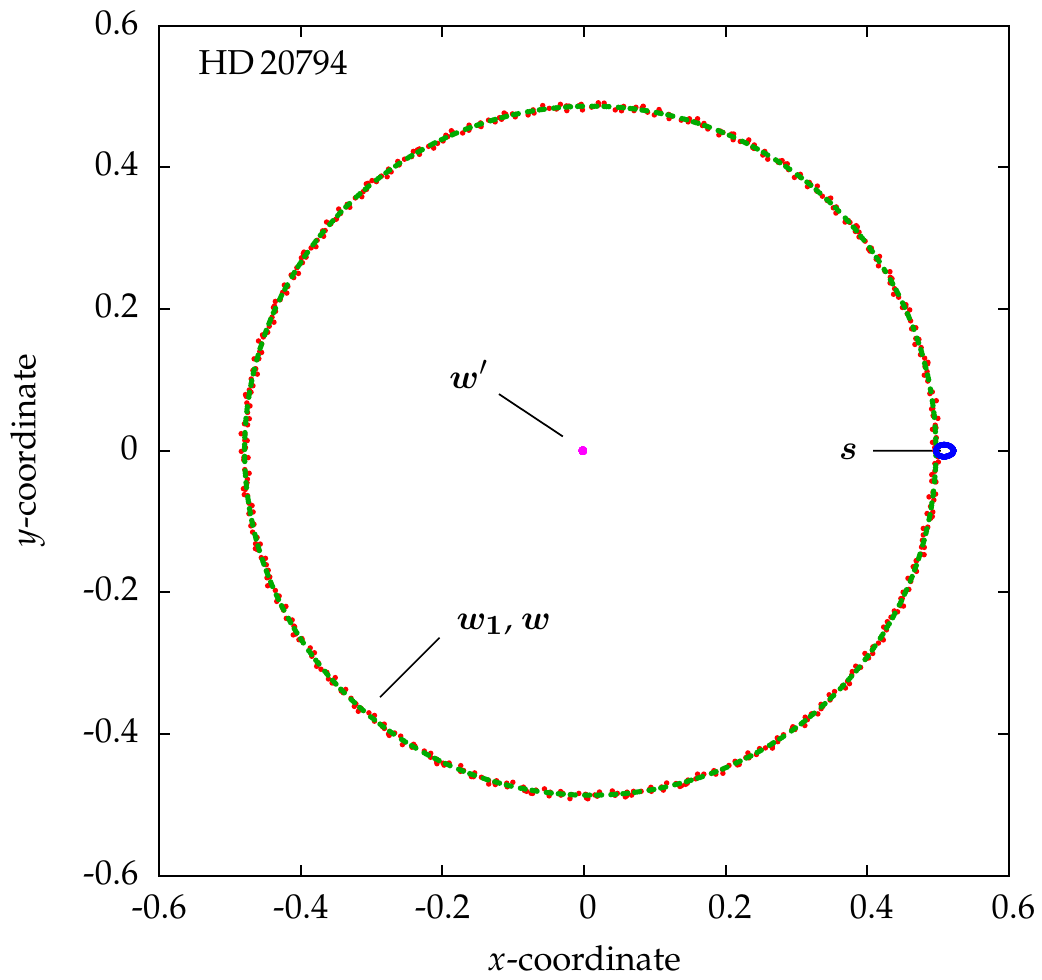}
\caption{\label{fig.HDa}Trajectory of the unit angular
momenta of the system HD\,20794 to which a giant planet with mass
$m'=1M_J$, semimajor axis $a'=20$ au, eccentricity $e'=0.1$, and
inclination $I'=30^\circ$ has been added. The colors are the same as in
Fig.~\ref{fig.Kaibb}. The frame is rotating at the stellar precession
frequency. }
\end{figure}
}
\newcommand{\figHDb}{
\begin{figure}[t]
\plotone{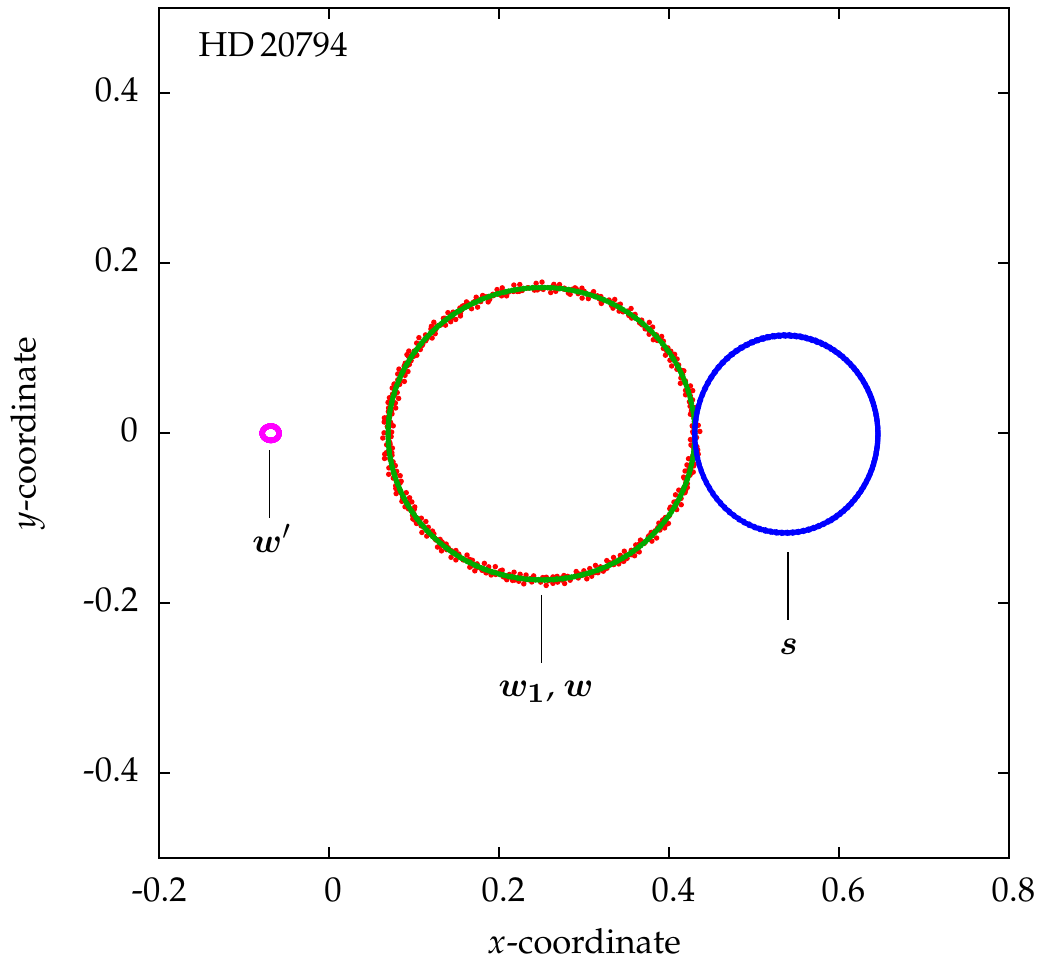}
\caption{\label{fig.HDb}Same as Fig.~\ref{fig.HDa} but with a lighter
perturbing planet $m'=0.04M_J$ comparable to Uranus.}
\end{figure}
}

\newcommand{\figBerlingot}{
\begin{figure}[t]
\plotone{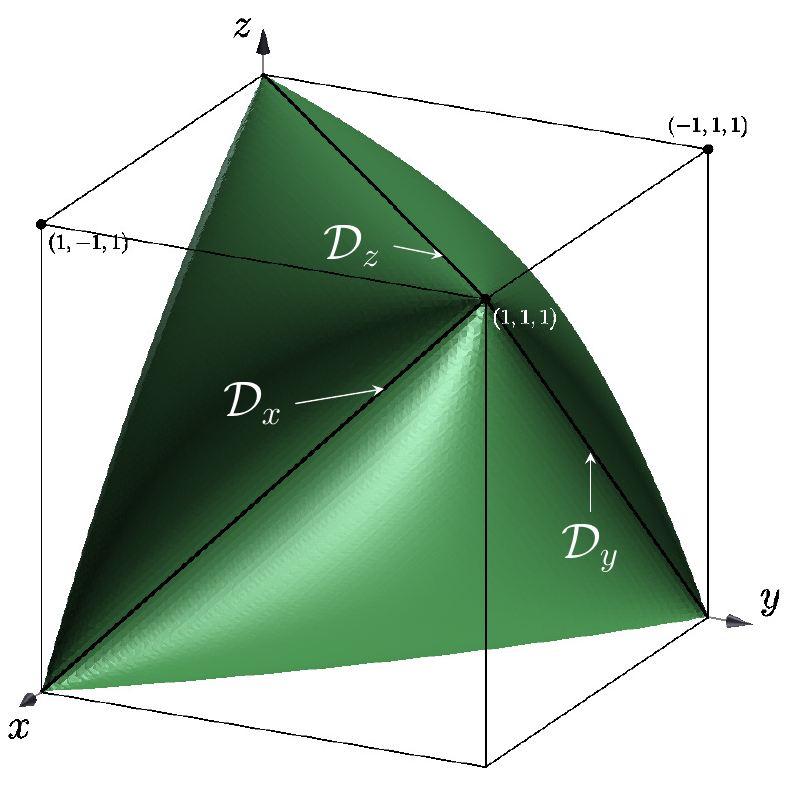}
\caption{\label{fig.Berl}Cassini Berlingot defined by $V^2(x,y,z)\geq 0$. As $V^2$ must be greater or equal to zero, the allowed region in the $(x,y,z)$ space is the interior of the Berlingot shape volume. The diagonals ${\cal D}_x$, ${\cal D}_y$, and ${\cal D}_z$ are also represented. See text for detail.}
\end{figure}
}
\newcommand{\figTypeSol}{
\begin{figure}[t]
\plotone{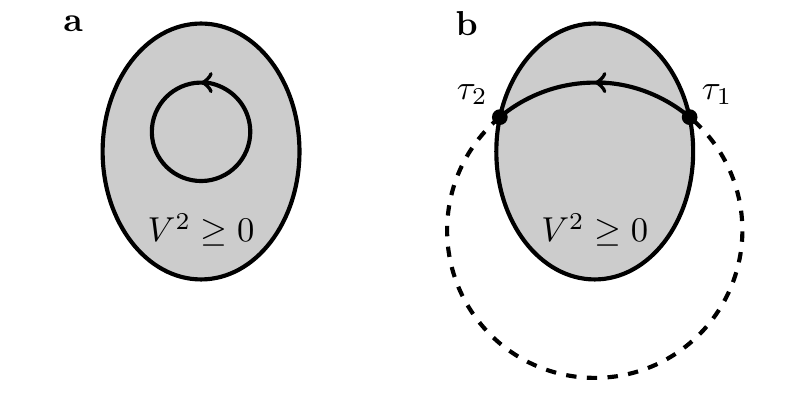}
\caption{\label{fig.TypeSol}Two types of relative motion. The shaded area represents the interior of the Berlingot ($V^2\geq 0$) and the circles are two different elliptic orbits. {\bf (a)} The orbit is fully inside the Berlingot. This configuration is called {\em special solution}. {\bf (b)} The orbit intersects the surface of the Berlingot in $\tau_1$ and $\tau_2$. This case is named {\em general solution}. The dashed section of the orbit (b) is forbidden because $V^2$ would be negative.}
\end{figure}
}
\newcommand{\figEquil}{
\begin{figure}[t]
\plotone{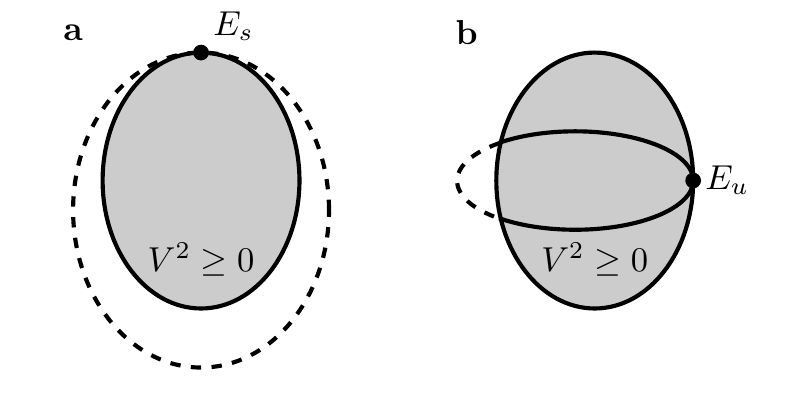}
\caption{\label{fig.equil}Two types of Equilibrium. The shaded areas represent the interior of the Berlingot ($V^2\geq 0$) and the ellipses are two different elliptic orbits. The dashed part of each trajectory is inaccessible. Equilibrium states are located at the points of tangency between elliptic trajectories and the surface of the Berlingot. {\bf (a)} When the tangency is from the outside of the Berlingot, the fixed point is stable ($E_s$). {\bf (b)} When the tangency is from the inside, the fixed point is unstable ($E_u$).}
\end{figure}
}
\newcommand{\figQuadric}{
\begin{figure}[t]
\plotone{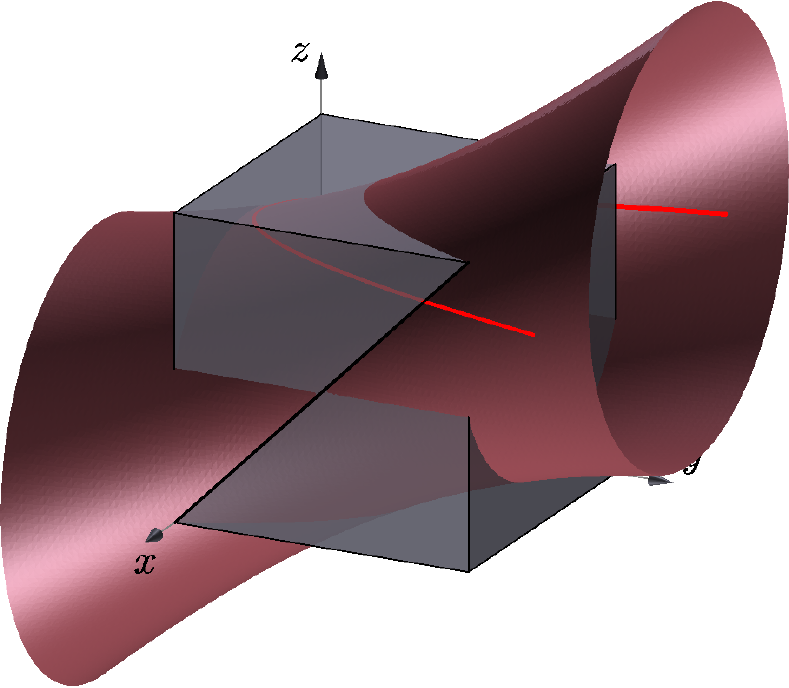}
\caption{\label{fig.quadric}Quadric ${\cal S}$ (red) defined as the
union of all elliptic trajectories starting at $x(0)=1$ and $y(0)=z(0)$
(thick black diagonal). The red curve is an example of elliptic trajectory 
with initial condition $y(0)=z(0)=\cos^{-1}45^\circ$. The cube ${\cal C}$ is
represented in gray.}
\end{figure}
}
\newcommand{\figSB}{
\begin{figure*}
\begin{center}
\includegraphics[width=0.8\linewidth]{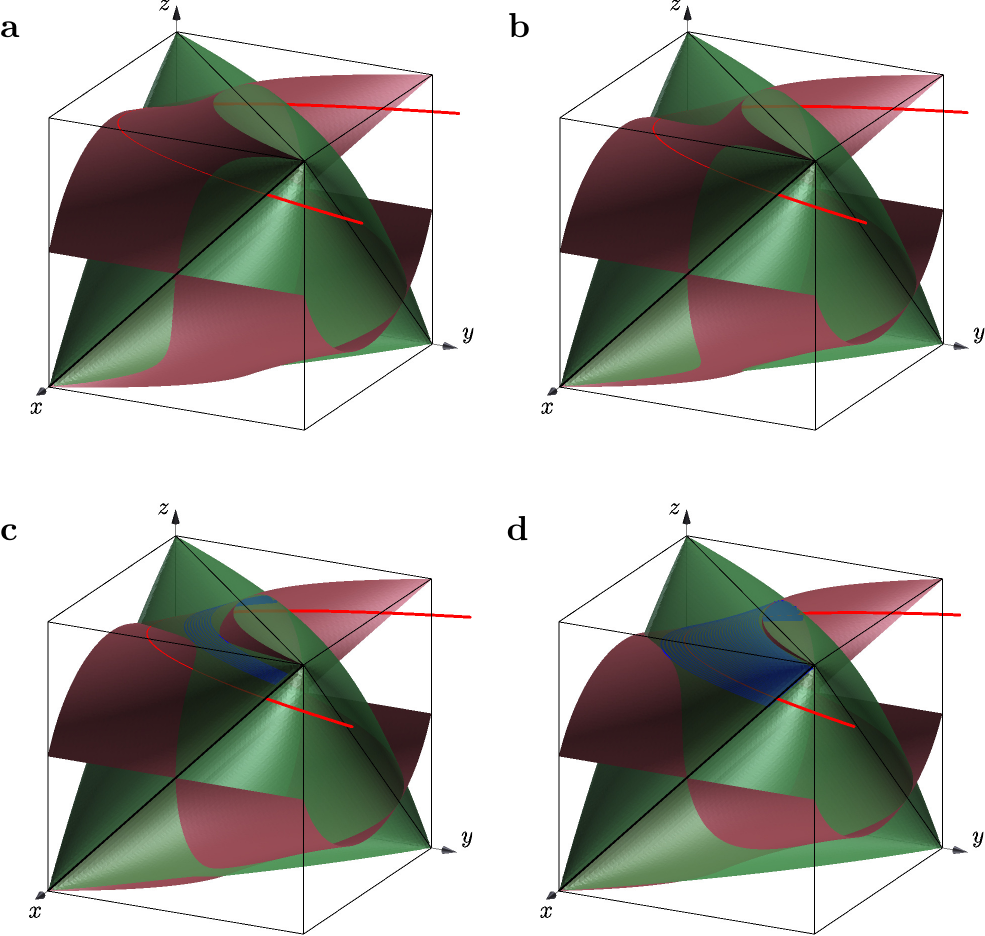}
\caption{\label{fig.SB}Modification of the topology of the spin-orbit
evolution in the 55 Cancri system as the companion's orbit shrinks. Parameters
are those of Fig.~\ref{fig.Kaibb} except the companion semiminor axis which
takes the values $b'=$190, 182, 180, and 170 in the panels {\bf a}, {\bf b},
{\bf c}, and {\bf d}, respectively. As $b'$ decreases from panel {\bf a} to {\bf d}, the quadric surface (violet) also shrinks. In panels {\bf c} and {\bf d}, a gap is open allowing some trajectories (blue) to reach low values of $x$ -- and thus high spin-orbit misalignment -- while remaining inside the Berlingot (green). The elliptic orbit of Fig.~\ref{fig.quadric} is still represented in red.}
\end{center}
\end{figure*}
}
\newcommand{\figMaxTilta}{
\begin{figure*}[t!]
\centering
\includegraphics[width=0.8\linewidth]{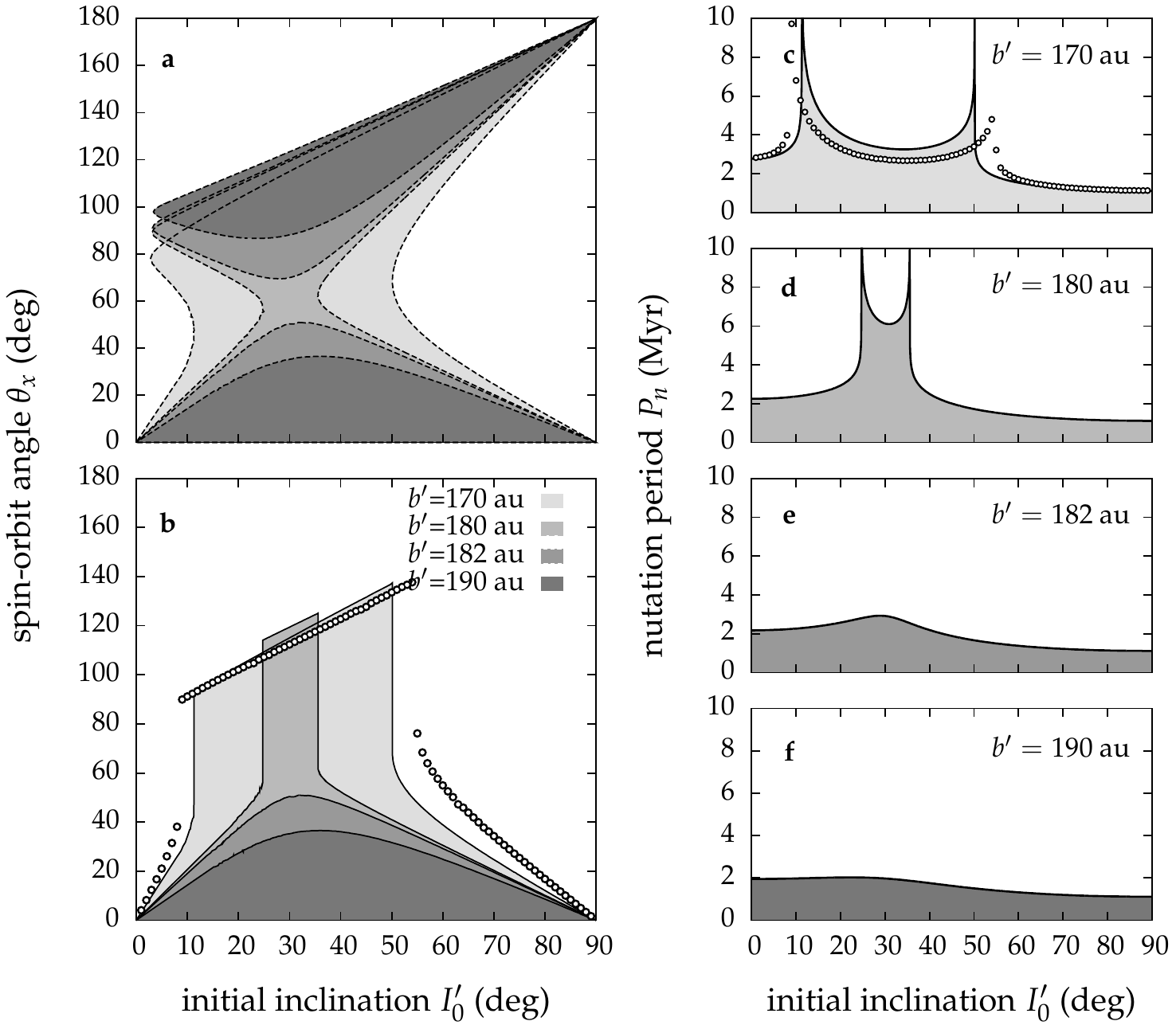}
\caption{\label{fig.maxtilta} {\bf a} Intersections ${\cal S}\cap{\cal
B}$ seen in the plane $(I'_0,\theta_x)$. {\bf b} Range of accessible
spin-orbit angles $\theta_x$ with initial condition $\theta_x(0)=0$
($x$-axis). {\bf c-f} Nutation periods computed from Eq.~(\ref{eq.Pn}).
Parameters are those of the 55~Cancri system (same as in
Fig.~\ref{fig.SB}), colors from dark to light correspond to $b'=$190,
182, 180, and 170 au, respectively. Points in the panels {\bf b}
and {\bf c} are the results of numerical integrations with $b'=170$ au.}
\end{figure*}
}
\newcommand{\figMaxTiltb}{
\begin{figure*}[t!]
\centering
\includegraphics[width=0.8\linewidth]{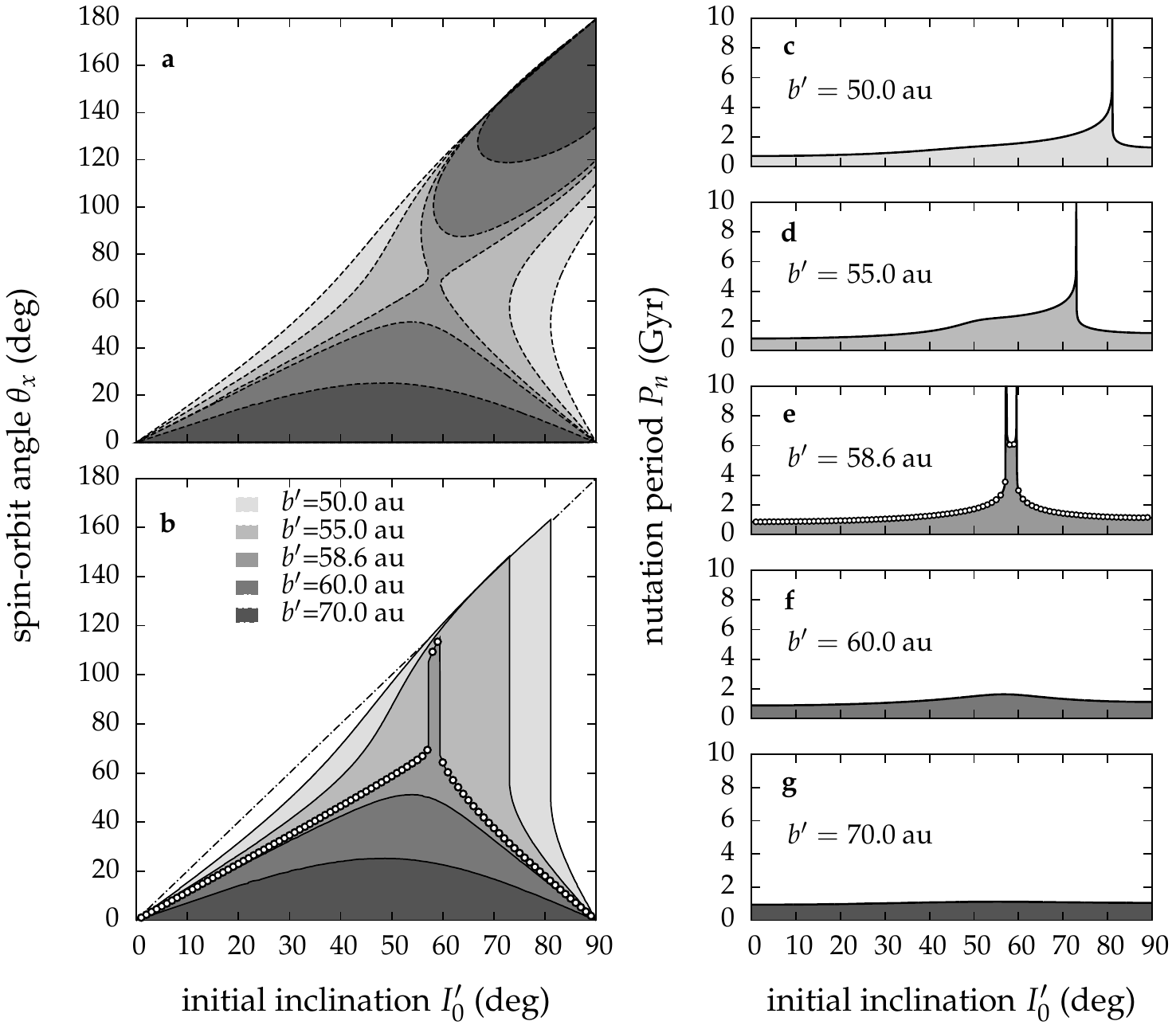}
\caption{\label{fig.maxtiltb} Same as Fig.~\ref{fig.maxtilta} with the
parameters of HD~20794, $m'= 1M_{\rm J}$, and $b'$=70, 60, 58.6, 55, and
50 au from dark to light gray, respectively. A series of numerical simulations
with $b'=58.6$ au are plotted in panels {\bf b} and {\bf e}, showing 
that the analytic model performs excellently for this system.  The dash-dotted line in
panel {\bf b} corresponds to $\theta_x = 2I'_0$.}
\end{figure*}
}
\newcommand{\figRegimes}{
\begin{figure*}[t]
\centering
\includegraphics[width=\linewidth]{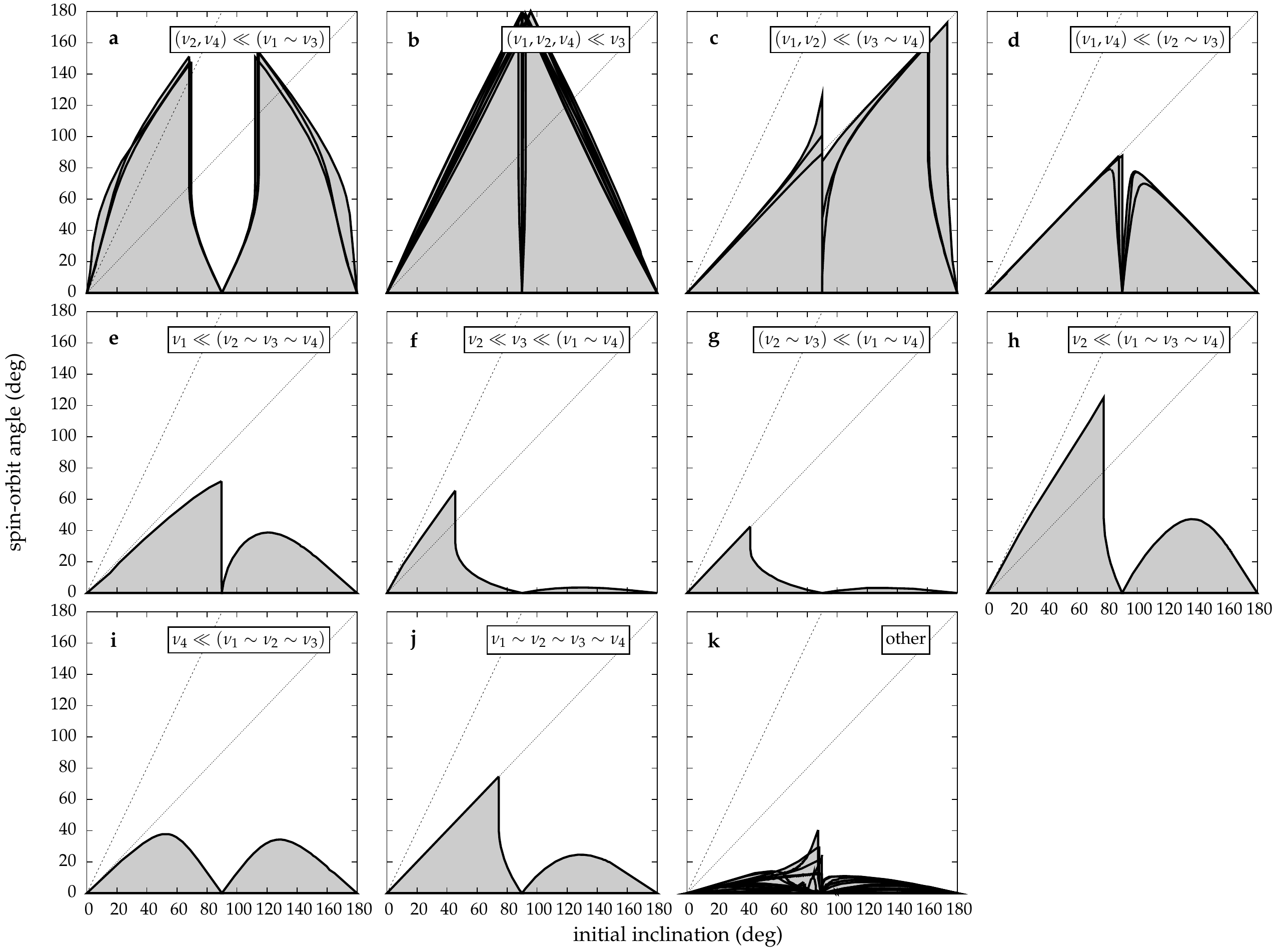}
\caption{\label{fig.regimes} Nutation amplitude $\theta_{x,\rm max}$ as
a function of the initial inclination $I'_0$ of the perturber in
different regimes. Similar behaviors are grouped together in same
panels. {\bf a} Cassini regime. {\bf b-c} Pure orbital regime. 
{\bf d-e} Laplace regime. {\bf f-j} Hybrid regime. {\bf k} All remaining
configurations leading to low nutation amplitudes. The dashed line and
the dotted line represent $\theta_{x,\rm max}=2 I'_0$ and $\theta_{x,\rm
max} = I'_0$, respectively.}
\end{figure*}
}
\newcommand{\figCassini}{
\begin{figure}[t!]
\plotone{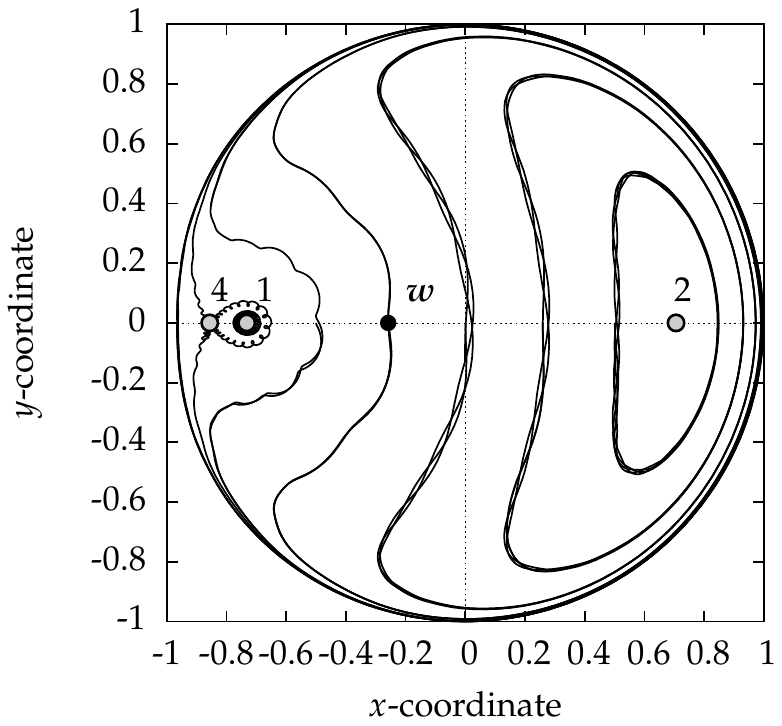}
\caption{\label{fig.cassini} Evolution of $\vec s$ obtained by numerical
integration of the 55~Cnc system with
$m'=0.26M_\odot$, $b'=170$ au, and $I'_0=15^\circ$. Trajectories are
plotted in the invariant plane in a frame rotating with the precession
period of the system. The Cassini states are labeled 1, 2, and
4. The point labeled $\vec w$ represents the position of $\vec w$ which
evolves very slightly in this frame.}
\end{figure}
}
\newcommand{\figValidity}{
\begin{figure}[t!]
\plotone{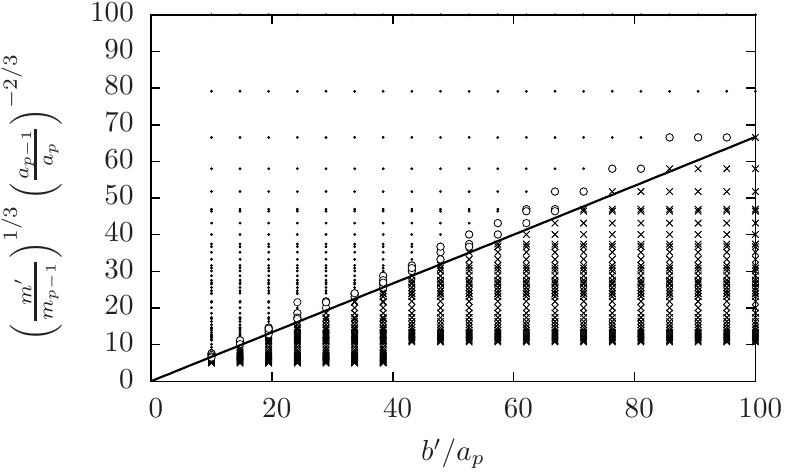}
\plotone{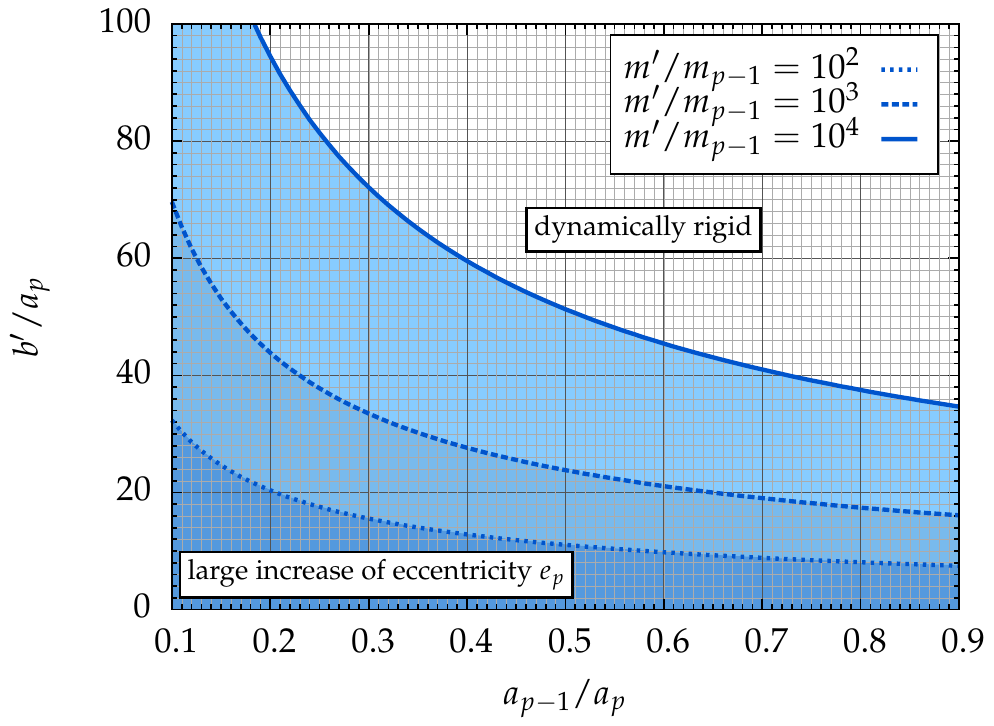}
\caption{\label{fig.validity} Determination of the validity of the analytical model. Top: results of the 2\,400 simulations done with $I'_0=89^\circ$. Dots, crosses, and open circles represent solutions with $\mathrm{max}(e_p)$ greater than 0.8, lower than 0.01, and intermediate, respectively. The solid line is the threshold (\ref{eq.validity}). Bottom: shaded areas are regions of high eccentricity increase as delimited by Eq.~(\ref{eq.validity}). The criterion is given for different $m'/m_{p-1}$ (mass of the companion / mass of the penultimate planet) in terms of the two outermost planet semimajor axis ratio $(a_{p-1}/a_p)$ versus the ratio $b'/a_p$ (companion semiminor axis / outer planet semimajor axis).}
\end{figure}
}
\newcommand{\TabCnc}{
\begin{table}[t]
\caption{\label{tab.Cnc}Masses and orbital elements of the planets of
the 55 Cnc system.}
\begin{tabular}{lrrrrrr} \hline \hline
Planet       & \multicolumn{1}{c}{$m$}      & 
               \multicolumn{1}{c}{$a$}      & 
               \multicolumn{1}{c}{$e$}      & 
               \multicolumn{1}{c}{$\varpi$} & 
               \multicolumn{1}{c}{$I$}      &
               \multicolumn{1}{c}{$\Omega$} \\
       & ($M_J$) & (au)  &       & ($^\circ$)    & ($^\circ$) & ($^\circ$) \\ \hline
e      & 0.025   & 0.015 & 0.00  &     0         &   0.70     &      0     \\
b      & 0.788   & 0.113 & 0.01  &   147         &   1.00     &      0     \\
c      & 0.164   & 0.237 & 0.06  &    99         &   0.30     &      0     \\
f      & 0.143   & 0.771 & 0.13  &   180         &   0.00     &      0     \\
d      & 3.660   & 5.700 & 0.03  &   189         &   0.03     &    180     \\ \hline
\end{tabular} \\[0.2em]
{\small {\bf notes.} Data are those of \citet{Dawson_Fabrycky_ApJ_2010}, except
semimajor axes which have been computed for a central mass
$m_0=0.905M_\odot$ instead of $0.94M_\odot$, and initial
inclinations which are introduced to avoid coplanar evolutions.\\}
\end{table}
}
\newcommand{\TabHD}{
\begin{table}
\caption{\label{tab.HD}Masses and orbital elements of the planets of
the HD~20794 system.}
\begin{tabular}{lrrrrrr} \hline \hline
Planet       & \multicolumn{1}{c}{$m$}      & 
               \multicolumn{1}{c}{$a$}      & 
               \multicolumn{1}{c}{$e$}      & 
               \multicolumn{1}{c}{$\varpi$} & 
               \multicolumn{1}{c}{$I$}      &
               \multicolumn{1}{c}{$\Omega$} \\
       & ($M_\oplus$) & (au)  &   & ($^\circ$)    & ($^\circ$) & ($^\circ$) \\ \hline
b      & 3.28    & 0.1288 & 0.00  &     0         &   0.10     &      0     \\
c      & 2.91    & 0.2172 & 0.01  &    90         &   0.30     &      0     \\
d      & 5.83    & 0.3733 & 0.03  &   270         &   0.15     &    180     \\ \hline
\end{tabular} \\[0.2em]
{\small {\bf notes.} Data are derived from \citet{Pepe_etal_AA_2011},
masses and semimajor axis have computed for a central mass
$m_0=0.85M_\odot$ instead of $0.70M_\odot$. The eccentricity and the
inclination are introduced solely to avoid circular and coplanar
evolutions.\\}
\end{table}
}
\newcommand{\Tabnotation}{
\begin{table*}
\begin{center}
\caption{\label{tab.notation}Notation.}
{\small
\begin{tabular}{cllll} \hline
& \multicolumn{2}{l}{variable} & Ref. & description \\ \hline \hline
\multirow{5}{*}{\rotatebox{90}{Hamiltonian\ }}
& \mtwo{$H_{\rm tot}$} & Eq.~(\ref{eq.Htot})& secular Hamiltonian of the numerical system \\
& \mtwo{$H$} & Eqs.~(\ref{eq.Hbar}, \ref{eq.Hxyz}) & secular Hamiltonian of the analytical model \\
& \mtwo{$K$} & Eq.~(\ref{eq.Kxyz}) & first integral of the analytical problem \\
& \mtwo{$\alpha$, $\gamma$} & Eq.~(\ref{eq.alpbetgam}) & coefficients of the analytical Hamiltonian $H$: $\sum \alpha_k$ and $\sum \gamma_k$, respectively \\
& \mtwo{$\lap{s}{k}$} & & Laplace coefficient \\
\hline
\multirow{7}{*}{\rotatebox{90}{timescales}}
& \mtwo{$\nu_1$} & Eq.~(\ref{eq.nus}) & precession frequency $\alpha/L$      of $\vec s$ relative to $\vec w$ \\
& \mtwo{$\nu_2$} & Eq.~(\ref{eq.nus}) & precession frequency $\alpha/G$      of $\vec w$ relative to $\vec s$ \\
& \mtwo{$\nu_3$} & Eq.~(\ref{eq.nus}) & precession frequency $\gamma/G$      of $\vec w$ relative to $\vec w'$ \\
& \mtwo{$\nu_4$} & Eq.~(\ref{eq.nus}) & precession frequency $\gamma/G'$ of $\vec w'$ relative to $\vec w$ \\
& \mtwo{$\nu_a$, $\nu_b$, $\nu_c$, $\nu_d$} && permutation of $\nu_1$, $\nu_2$, $\nu_3$, $\nu_4$ \\
& \mtwo{$P_{\rm nut}$} & Eq.~(\ref{eq.Pn}) & nutation period \\
& \mtwo{$P_{\rm prec}$} & Eq.~(\ref{eq.Pp}) & precession period \\
\hline
\multirow{11}{*}{\rotatebox{90}{stellar parameters}}
& \mtwo{$m_0$} & & mass \\
& \mtwo{$R_0$} & & radius \\
& \mtwo{$J_2$} & Eq.~(\ref{eq.Jom}) & quadrupole gravitational harmonic \\
& \mtwo{$k_2$} & & second fluid Love number \\
& \mtwo{$C$} & & moment of inertia along the short axis \\
& \mtwo{$\omega_0$} & & rotation speed \\
& \mtwo{$P_0$} & & rotation period $2\pi/\omega_0$ \\
& \mtwo{$\epsilon$} & & obliquity relative to the reference plane \\
& \mtwo{$\psi$} & & precession angle \\
& \mtwo{$\vec s$} & & stellar spin axis \\
& \mtwo{$\vec L$} & & stellar angular momentum $C\omega_0\vec s$\\
\hline
\multirow{11}{*}{\rotatebox{90}{$j$th planet and companion}}
& $m_j$ & $m'$ & & mass \\ 
& $a_j$ & $a'$ & & semimajor axis \\
& & $b'$ & & semiminor axis $a'(1-e'^2)^{1/2}$ \\
& $P_j$ & $P'$ & & revolution period \\
& $e_j$ & $e'$ & & eccentricity \\
& $I_j$ & $I'$ & & absolute inclination (with respect to the reference plane) \\
& $\Omega_j$ & $\Omega'$ & & longitude of the ascending node \\
& $\vec e_j$ & & & eccentricity vector \\
& $\vec j_j$ & & & dimensionless orbital angular momentum $(1-e_j^2)^{1/2}\vec w_j$ \\
& $\vec w_j$ & $\vec w'$ & & unit orbital angular momentum \\
& $\vec G_j$ & $\vec G'$ & & orbital angular momentum \\
\hline
\multirow{7}{*}{\rotatebox{90}{other variables}}
& \mtwo{$\vec w$} & & unit vector of $\vec G$ \\
& \mtwo{$\vec G$} & Eq.~(\ref{eq.vecG}) & total angular momentum of the planet system $G\vec w = \sum G_j \vec w_j$ \\
& \mtwo{$\vec W$} & Eq.~(\ref{eq.Wtot}) & total angular momentum $\vec L + \vec G + \vec G'$ \\
& \mtwo{$x=\cos\theta_x$} & Eq.~(\ref{eq.xyz}) & cosine of the stellar obliquity relative to the planets plane $\vec s\cdot\vec w$ \\
& \mtwo{$y=\cos\theta_y$} & Eq.~(\ref{eq.xyz}) & cosine of the stellar obliquity relative to the companion's orbit $\vec s\cdot\vec w'$ \\
& \mtwo{$z=\cos\theta_z$} & Eq.~(\ref{eq.xyz}) & cosine of the mutual inclination between the planets and the companion $\vec w\cdot\vec w'$ \\
& \mtwo{$\tau$} & Eq.~(\ref{eq.taut}) & fictitious time used to parametrize elliptic orbits in $(x,y,z)$ \\
\hline
\multirow{8}{*}{\rotatebox{90}{geometric objects\ }}
& \mtwo{${\cal E}$} & & elliptic orbit in $(x,y,z)$ satisfying $H(x,y,z)=h$ and $K(x,y,z)=k$ for two reals $(h,k)$ \\
& \mtwo{$\cal C$, $\partial \cal C$} & & cube $[-1,1]\times[-1,1]\times[-1,1]$ in $(x,y,z)$ and its boundary, respectively \\
& \mtwo{$\cal B$, $\partial\cal B$} & Fig.~\ref{fig.Berl} & Cassini Berlingot defined by $V^2(x,y,z)\geq 0$ and its
boundary, respectively \\
& \mtwo{${\cal D}_x$, ${\cal D}_y$, ${\cal D}_z$} & Fig.~\ref{fig.Berl} & diagonals which are the intersections of $\partial \cal C$ and $\partial \cal B$ \\
& \mtwo{$\cal S$} & Fig.~\ref{fig.quadric} & hyperbolic surface equal to the union of all elliptic
trajectories intersecting ${\cal D}_x$ \\
& \mtwo{$V$} & Eqs.~(\ref{eq.Vswn}, \ref{eq.Vxyz}) & oriented volume generated by $(\vec s, \vec w, \vec w')$ \\
& \mtwo{$S$} & Eq.~(\ref{eq.Sxyz}) & quadric function defining the surface $\cal S$ \\
& \mtwo{$A_x$, $A_z$} & Eq.~(\ref{eq.AxAz}) & length scales of the elliptic orbit in $(x,y,z)$ \\
\hline
& \mtwo{$\cal G$} & & gravitational constant \\
& \mtwo{$c$} & & speed of light \\
\hline
\end{tabular}
}
\end{center}
\end{table*}
}
\shorttitle{Perturbed compact planetary systems}
\shortauthors{Bou\'e G. \& Fabrycky D.}
\begin{document}
\title{Compact planetary systems perturbed by an inclined companion: II. Stellar spin-orbit evolution}
\author{Gwena\"el Bou\'e\altaffilmark{1,2} and Daniel C. Fabrycky\altaffilmark{1}}
\altaffiltext{1}{Department of Astronomy and Astrophysics, University of Chicago,
5640 South Ellis Avenue, Chicago, IL 60637, USA}
\altaffiltext{2}{Astronomie et Syst\`emes Dynamiques, IMCCE-CNRS UMR 8028,
Observatoire de Paris, UPMC, 77 Av. Denfert-Rochereau, 75014 Paris, France.}
\email{boue@uchicago.edu}
\begin{abstract}
The stellar spin orientation relative to the orbital planes of multiplanet systems are becoming accessible to observations.  Here, we analyze and classify different types of spin-orbit evolution in compact multiplanet systems perturbed by an inclined outer companion.  Our study is based on classical secular theory, using a vectorial approach developed in a separate paper.  When planet-planet perturbations are truncated at the second order in eccentricity and mutual inclination, and the planet-companion perturbations are developed at the quadrupole order, the problem becomes integrable.  The motion is composed of a uniform precession of the whole system around the total angular momentum, and in the rotating frame, the evolution is periodic. Here, we focus on the relative motion associated to the oscillations of the inclination between the planet system and the outer orbit, and of the obliquities of the star with respect to the two orbital planes. The solution is obtained using a powerful geometric method. With this technique, we identify four different regimes characterized by the nutation amplitude of the stellar spin-axis relative to the orbital plane of the planets. In particular, the obliquity of the star reaches its maximum when the system is in the Cassini regime where planets have more angular momentum than the star, and where the precession rate of the star is similar to that of the planets induced by the companion. In that case, spin-orbit oscillations exceed twice the inclination between the planets and the companion.  Even if mutual inclination is only $\simeq 20^\circ$, this resonant case can cause the spin-orbit angle to oscillate between perfectly aligned and retrograde values. 
\end{abstract}
\section{Introduction}
Hot or eccentric Jupiters only constitute a small fraction of the exoplanets discovered to date. Among those with short orbital periods, most are smaller, less massive, and part of compact multiplanet systems \citep{Howard_etal_Science_2010, Howard_etal_ApJL_2012, Howard_Science_2013, Petigura_etal_ApJ_2013}. Due to the smaller planetary radii and larger orbital periods, the efficiency of the standard method to measure spin-orbit angle in systems with hot Jupiters, based on the Rossiter-McLaughlin effect \citep{Holt_AA_1893, Rossiter_ApJ_1924, McLaughlin_ApJ_1924}, decreases significantly in multiplanet systems. Only two multiplanet systems, called KOI-94 and Kepler-25, have been studied with this technique \citep{Hirano_etal_ApJL_2012, Albrecht_etal_ApJ_2013}. Two other methods have been implemented to measure the spin-orbit angle in multiplanet systems, the stellar spot crossing technique on Kepler-30 \citep{Sanchis-Ojeda_etal_nature_2012}, and asteroseismology on Kepler-50 and Kepler-65 \citep{Chaplin_etal_ApJ_2013}.  The five systems prove to be compatible with perfect spin-orbit alignments, {\em a priori} suggesting that multiplanet systems are preferentially in the equatorial plane of their star. A sixth system, Kepler-56 \citep{Huber_etal_Science_2013} shows coplanarity between two transiting planets, but misalignment to the star; a distant companion, detected by radial velocity, may be responsible. Therefore we are led to ask what happens to the stellar obliquity if a coplanar multiplanet system is accompanied by a distant planetary companion or is embedded in a binary stellar system? 
Dynamically, large stellar obliquity in isolated close-in planet system can be the outcome of either planet-planet scattering \citep{Nagasawa_etal_ApJ_2008,Beauge_Nesvorny_ApJ_2012} or Lidov-Kozai excitation by an outer inclined perturber \citep{Fabrycky_Tremaine_ApJ_2007, Correia_etal_CeMDA_2011, Naoz_etal_nature_2011, Naoz_etal_ApJL_2012}. Moreover, if the inner eccentricity is large, even a coplanar outer object can flip the planet's orbit by 180$^\circ$ \citep{Li_etal_ApJ_2014}. More generally, in single as well as in multiplanetary systems, spin-orbit misalignment may also result from the magnetic interaction between the protostar and its circumstellar disc \citep{Lai_etal_MNRAS_2011} or from the solid precession of the protoplanetary disc induced by an inclined companion \citep{Batygin_nature_2012, Batygin_Adams_ApJ_2013, Lai_MNRAS_2014}.
In multiplanet systems surrounded by an outer stellar companion, apsidal precession frequencies are dictated by the companion {\em and} by the planet-planet interactions. As a consequence, even at high inclination, if the planet system is sufficiently packed, planet-planet interactions dominate the apsidal motion, the evolution is stabilized with respect to the Lidov-Kozai mechanism, eccentricities remain small, and all planets move in concert \citep{Innanen_etal_AJ_1997, Takeda_etal_ApJ_2008, Saleh_Rasio_ApJ_2009}. These systems are classified as dynamically rigid\footnote{Note that our definition of dynamically rigid is more stringent than that of \citet{Takeda_etal_ApJ_2008} who also include the case where planet eccentricities increase in concert.}. Although the Lidov-Kozai evolution is quenched, the planetary mean plane still precesses if it is inclined relative to the orbit of the companion, and can eventually lead to spin-orbit misalignment with the central star.  \citet{Kaib_etal_ApJ_2011} applied this idea to the 55~Cancri multiplanet system which has a stellar companion, and concluded that the planets are likely misaligned with respect to the stellar equator.  However, the results only hold as long as the stellar spin-axis is weakly coupled to the planets orbit.  We show here that this condition is not satisfied for the 55~Cancri system unless the semiminor axis of the perturber is very small, of the order of 180 au (periastron distance $\lesssim 30$ au), whereas the projected separation is 1065 au \citep{Mugrauer_etal_AN_2006}.  Nevertheless, the required conditions for this mechanism may have been met earlier in the history of this system in particular, or other systems in general. Indeed, in our own solar system, for instance, the Sun is weakly coupled to the ecliptic and its obliquity of 7$^\circ$ might be the signature of an earlier tilt of the planet system \citep{Tremaine_Icarus_1991}. Moreover, analyzing a similar problem where a protoplanetary disk takes the place of the compact planet system, \citet{Batygin_nature_2012} showed that this mechanism is able to tilt forming planetary systems around slow rotator T Tauri pre-main sequence stars.
Here, we revisit the problem composed of a dynamically rigid system perturbed by a stellar or a planetary companion on a wide and inclined orbit. The inner planets are assumed to have low eccentricities and mutual inclinations comparable to or lower than those of our own solar system. According to these assumptions, orbital evolution induced by tides is expected to be weak and is neglected. These hypotheses are motivated by statistical studies of compact exoplanet systems detected by {\em Kepler} or by radial velocity \citep[e.g.,][]{Tremaine_Dong_AJ_2012, Figueira_etal_AA_2012, Fabrycky_etal_ApJ_2013, Wu_Lithwick_ApJ_2013}. However, we allow the overall plane of planets to tilt by an arbitrary angle. A hierarchical companion is included, which is allowed to have any eccentricity and inclination.  The main goal of this study is to follow the evolution of the inclination of the planet system with respect to the spin-axis of the parent star. Thus, the interaction between the stellar spin-axis and the orbital motion of the inner planets is taken into account. 
For this study, we exploit the results of the so-called ``3-vector problem'' which has been solved geometrically in \citep[hereafter BL06]{Boue_Laskar_Icarus_2006} and in \citep[hereafter BL09]{Boue_Laskar_Icarus_2009}. The 3-vector problem aims to model the secular evolution of three coupled angular motions such as the lunar problem with the planet spin and the orbital angular momenta of the satellite and the star \citepalias{Boue_Laskar_Icarus_2006} or the binary asteroid problem with two spin-axes and their mutual orbital motion \citepalias{Boue_Laskar_Icarus_2009}. Here, the three vectors are the spin of the star, the total orbital angular momentum of the planet system, and that of the companion. In Section~\ref{sec.integrable}, we recall the main results of the 3-vector problem, and we also provide a new integral expression for the precession frequency. Then, in Section~\ref{sec.nbodyproblem} we employ the vectorial formalism of the classical secular theory that we described in a previous paper (Boue and Fabrycky 2014; BF14) and we show how the three vector problem emerges from this general secular model. The validity of the simplification is also discussed. In this work, we thus consider two different models which correspond to two levels of approximation. On the one hand, the perturbing function is expanded at the fourth order in planet eccentricity and mutual inclination and at the octupole in the interaction between each planet and the companion. This model provides accurate results but is non-integrable and has to be solved numerically. On the other hand, the system is described by the integrable three vector problem which gives deeper geometrical insight. In the following, we refer to the former as the {\em numerical} model and to the latter as the {\em analytical} model.
 In Section~\ref{sec.numerical}, the two models are compared in their application to real exoplanet systems. Then, we exploit more deeply the possibilities of the analytical model to span the parameter space and identify four different regimes of evolution in Section~\ref{sec.global}. The conclusions are given in the last section.
\Tabnotation
\section{Three-vector problem}
\label{sec.integrable}
This section summarizes a few key results associated to the so-called ``3-vector problem'' described in \citepalias{Boue_Laskar_Icarus_2006, Boue_Laskar_Icarus_2009}. In the context of this paper, the three vectors are the angular momentum of the star $\vec L=L\vec s$, the orbital angular momenta of the planet system $\vec G=G\vec w$, and that of the companion $\vec G' = G'\vec w'$, where $\vec s$, $\vec w$, and $\vec w'$ are unit vectors. The 3-vector problem assumes that the evolution is governed by a Hamiltonian of the form
\be
H = -\frac{\alpha}{2}(\vec s\cdot\vec w)^2 - \frac{\beta}{2}(\vec s\cdot\vec w')^2 -
     \frac{\gamma}{2}(\vec w\cdot\vec w')^2\ ,
\label{eq.Hbar}
\ee
where $\alpha$, $\beta$, and $\gamma$ are constant parameters representing the coupling between the planetary system and both the stellar rotation and the binary orbit, respectively. Their expression will be derived in the subsequent section. Note that in contrast to the more general 3-vector problem, here we neglect the direct interaction between the stellar spin and the orbit of the companion, i.e., we set $\beta=0$. The equations of motion are
\be
\EQM{
\frac{d\vec s}{dt} &= 
  -\frac{1}{L}\vec s \times \grad{\vec s}{H} \ ;\crm
\frac{d\vec w}{dt} &= 
  -\frac{1}{G} \vec w \times \grad{\vec w}{H}\ ;\crm
\frac{d\vec w'}{dt} &= 
  -\frac{1}{G'} \vec w' \times \grad{\vec w'}{H}\ ,
}
\ee
which leads to
\be
\EQM{
\frac{d\vec s}{dt} &=&
  -\frac{\alpha}{L}(\vec s\cdot \vec w)\vec w\times \vec s\ ;\crm
\frac{d\vec w}{dt} &=&
  -\frac{\alpha}{G}(\vec s\cdot \vec w)\vec s\times \vec w
  -\frac{\gamma}{G}(\vec w'\cdot \vec w)\vec w'\times \vec w\ ;\crm
\frac{d\vec w'}{dt} &=&
  -\frac{\gamma}{G'}(\vec w'\cdot \vec w)\vec w\times \vec w'\ .
}
\label{eq.evolswn}
\ee
In prevision of the subsequent analysis, we set
\be
\EQM{
\nu_1 & = \alpha/L \ ,\crm
\nu_2 & = \alpha/G \ ,\crm
\nu_3 & = \gamma/G \ ,\crm
\nu_4 & = \gamma/G'\ .
\label{eq.nu1234}
}
\ee
These quantities are important as they are the characteristic precession frequencies of $\vec s$ around $\vec w$, of $\vec w$ around $\vec s$ and $\vec w'$, and of $\vec w'$ around $\vec w$, respectively.

\subsection{Integrability}
\label{sec.integrability}
The 3-vector problem is integrable \citepalias{Boue_Laskar_Icarus_2006, Boue_Laskar_Icarus_2009}. Let 
\be
\vec W = L\vec s + G\vec w + G'\vec w'
\label{eq.Wtot}
\ee
be the total angular momentum of the system. The general solution is a uniform rotation of the three vectors around the total angular momentum combined with a periodic motion in the rotating frame \citepalias{Boue_Laskar_Icarus_2006, Boue_Laskar_Icarus_2009}. The evolution is thus characterized by two frequencies or periods. Hereafter, the uniform rotation is referred to as the precession motion with period $P_{\rm prec}$, and the periodic loops described in the rotating frame are equally qualified as nutation in reference to the Earth-Moon problem, or simply as the relative motion with period $P_{\rm nut}$.
The relative motion can be solved elegantly with geometric arguments \citepalias{Boue_Laskar_Icarus_2006, Boue_Laskar_Icarus_2009}. It is also very important for our study for two reasons: it enables 1) to check if any system can be misaligned, and 2) to evaluate the timescale of the secular spin-orbit evolution which can then be compared to the lifetime of the system. Next, we recall its solution and main properties as derived in \citepalias{Boue_Laskar_Icarus_2006, Boue_Laskar_Icarus_2009}. Then, we present a new integral expression of the precession period.
\subsection{Relative motion}
In order to get the relative evolution of the system described by the Hamiltonian (\ref{eq.Hbar}), we follow the same derivation as in \citetalias{Boue_Laskar_Icarus_2006} and \citetalias{Boue_Laskar_Icarus_2009}. We denote
\be
\EQM{
x &= \vec s\cdot \vec w\ , \crm
y &= \vec s\cdot \vec w'\ , \crm
z &= \vec w\cdot \vec w'\ .
}
\label{eq.xyz}
\ee
Sometimes, we will also use the corresponding angles defined by $x=\cos\theta_x$, $y=\cos\theta_y$, and $z=\cos\theta_z$. In this coordinate system, the Hamiltonian reads as
\be
H = -\frac{\alpha}{2}x^2 -\frac{\gamma}{2}z^2\ .
\label{eq.Hxyz}
\ee
The conservation of the norm of each angular momentum $L$, $G$, and $G'$, as well as the total angular angular momentum of the system $\vec W$, Eq.~(\ref{eq.Wtot}), lead to the second constant of the motion
\be
\EQM{
K &= \frac{\norm{\vec W}^2-L^2-G^2-G'^2}{2} \crm
  &= LGx + LG' y +GG' z\ .
}
\label{eq.Kxyz}
\ee
Each trajectory of the relative motion in the $(x,y,z)$ frame is at the intersection of a cylinder defined by $H(x,y,z)=h$ and a plane defined by $K(x,y,z)=k$, where $h$ and $k$ are two constants given by the initial conditions. Trajectories are thus subsets of ellipses defined by the values $h$ and $k$ of the two first integrals of the motion.  Hereafter, we denote them as ${\cal E} = \{(x,y,z)\in\mathbb{R}^3 \mid H(x,y,z)=h, K(x,y,z)=k\}$. These ellipses can be parametrized as follows
\be
\EQM{
x(\tau) &= A_x \cos\tau\ , \crm
z(\tau) &= A_z \sin \tau\ , \crm
y(\tau) &= \frac{1}{LG'}\big(k - LGx(\tau) - GG' z(\tau)\big)\ ,
}
\label{eq.paramEll}
\ee
where
\be
\EQM{
A_x &= \sqrt{\frac{-2h}{\alpha}}\ , \quad
A_z &= \sqrt{\frac{-2h}{\gamma}}\ .
}
\label{eq.AxAz}
\ee
The change of time $t\mapsto \tau$ leading to the parametrization (\ref{eq.paramEll}) will be made explicit in section \ref{sec.periods}.
In general, systems do not cover the full ellipses. Indeed, $x$, $y$, and $z$ are dot products of unit vectors and the evolution is restricted inside the cube ${\cal C} = \{(x,y,z)\in[-1,1]^3\}$. There is also a more stringent additional constraint \citepalias{Boue_Laskar_Icarus_2006}. Let
\be
V = \vec s\cdot(\vec w\times \vec w')\ .
\label{eq.Vswn}
\ee
V represents the oriented volume of the parallelepiped generated by the vectors $\vec s$, $\vec w$, and $\vec w'$. In terms of the dot products $x$, $y$, $z$, the square of the volume $V$ is given by the Gram determinant
\be
V^2 = \left|
\begin{matrix}
1 & x & y \cr
x & 1 & z \cr
y & z & 1
\end{matrix}
\right| = 1-x^2-y^2-z^2+2xyz\ .
\label{eq.Vxyz}
\ee
When the three vectors $\vec s$, $\vec w$, and $\vec w'$ are coplanar, $V^2(x,y,z)=0$. This is the equation of a cubic surface known as Cayley's nodal cubic. The restriction of this surface to the cube ${\cal C}$ is displayed in Fig.~\ref{fig.Berl}. In \citetalias{Boue_Laskar_Icarus_2009}, this restriction is called {\em Cassini Berlingot}. The word Berlingot comes after a french hard candy with a similar shape. The name Cassini has been added in reference to the `Cassini states' characterized by the coplanarity of the same three vectors as in this problem. Thus `Cassini states' are located at the surface $V^2(x,y,z)=0$. Because the cubic (\ref{eq.Vxyz}) represents the square of the volume $V$, it must be positive. As a consequence, the evolution of the system is restricted inside the Berlingot ${\cal B} = \{(x,y,z)\in{\cal C} \mid V^2(x,y,z)\geq 0\}$. Here, ${\cal B}$ denotes the inside of the Berlingot, and $\partial {\cal B}$ its surface.
\figBerlingot
In Fig.~{\ref{fig.Berl}}, important diagonals ${\cal D}_x$, ${\cal D}_y$, and ${\cal D}_z$ are represented. They all belong to the intersection $\partial {\cal B}\cap \partial {\cal C}$, in which the three vectors ($\vec s, \vec w, \vec w'$) lie in the same plane. The point $(1,1,1)$ corresponds to the configuration where all three vectors are aligned and in the same direction. In that case, the system is fully coplanar with only prograde orbits. At the point $(1,-1,-1)$, the three vectors are still collinear, but $\vec w'$ is pointing in the opposite direction as $\vec s$ and $\vec w$. In that case, the system is also coplanar, but the outer companion is on a retrograde orbit. Along the diagonal ${\cal D}_x$ joining these two points, the planet system remains in the equatorial plane of the star, and the companion's inclination $\theta_y=\theta_z$ increases from 0$^\circ$ at $(1,1,1)$ to 180$^\circ$ at $(1,-1,-1)$. In the same way, along the diagonal ${\cal D}_y$, the angle $\theta_y=\cos^{-1}(\vec s\cdot \vec w')$ is constant and equal to 0, thus the equator of the star remains in the plane of the outer body, while the planet system tilts from $0^\circ$ at $(1,1,1)$ to $180^\circ$ at $(-1,1,-1)$. Finally, along the diagonal ${\cal D}_z$, $\vec w\cdot \vec w' = \cos \theta_z = 1$, the planet system remains in the plane of the outer perturber and the stellar obliquity $\theta_x=\theta_y$ increases from $0^\circ$ at $(1,1,1)$ to $180^\circ$ at $(-1,-1,1)$.
Two classes of solutions, drawn schematically in Fig.~\ref{fig.TypeSol}, exist \citepalias{Boue_Laskar_Icarus_2006}. The first class comprises {\em special solutions}, so called because the three vectors $\vec s$, $\vec w$, and $\vec w'$ never get coplanar and remain always almost orthogonal to each other, which is not usual. In this case, the ellipse ${\cal E}$ is fully contained inside the Berlingot ${\cal B}$.  The solutions of the second class are called {\em general solutions}. They are such that $\vec s$, $\vec w$, and $\vec w'$ periodically lie in the same plane, which happens when ${\cal E}$ crosses the surface of the Berlingot $\partial{\cal B}$. In the following, we limit our analysis to systems belonging to the second class only. For a detailed description of the special solutions, see \citetalias{Boue_Laskar_Icarus_2006} and \citetalias{Boue_Laskar_Icarus_2009}.
\figTypeSol
As stated before, the relative motion of the three vectors is integrable. In particular, the trajectory of any general solution in the $(x,y,z)$ frame is a continuous piece of ellipse. More precisely, trajectories are connected sections of ${\cal E} \cap {\cal B}$ with extremities at the surface $\partial{\cal B}$ of the Berlingot. These elliptic sections are swept back and forth indefinitely, each round trip defining a nutation oscillation, unless the ellipse ${\cal E}$ is tangent to the Berlingot $\partial {\cal B}$, in which case the point of tangency is a fixed point of the relative motion \citepalias{Boue_Laskar_Icarus_2006}. The stability of each relative equilibrium is easily obtained as follows. If the ellipse is tangent to the Berlingot `from the outside' (Fig.~\ref{fig.equil}.{\bf a}), the trajectory is a singleton and thus the relative equilibrium is stable.  On the other hand, if the ellipse is tangent to the Berlingot `from the inside' (Fig.~\ref{fig.equil}.{\bf b}), the trajectory goes within the Berlingot, and the relative equilibrium is unstable. We stress again that because the fixed points are located at the surface ($V^2(x,y,z)=0$) of the Berlingot, in the equilibrium states the three vectors $\vec s$, $\vec w$, and $\vec w'$ lie in the same plane, as in the Cassini states. 
\figEquil
\subsection{Amplitudes of nutation}
\label{sec.amplitude_nutation}
The precession motion is a solid rotation of the three vectors $\vec s, \vec w, \vec w'$ leaving their mutual angles unchanged. Conversely, oscillations of inclinations and spin-orbit angles constitute the nutation motion. The amplitude of these angles are thus dictated by the relative motion taking place in the space $(x, y, z)$. More precisely, the extrema of the stellar obliquity $\theta_x$ with respect to the mean planet plane, of the stellar obliquity $\theta_y$ with respect to the outer orbit, and of the mutual inclination $\theta_z$ between the planet mean plane and the outer orbit correspond to the extrema of the dot products $x, y, z$ reached during one nutation period. In the following, most of the derivation is done for $\theta_x$, but the method is equivalent for $\theta_y$ and $\theta_z$.
With the parametrization (\ref{eq.paramEll}), the extrema of $x$ are attained when $dx/d\tau=0$ or when the three vectors lie in the same plane, i.e., $V^2(\tau) = V^2(x(\tau), y(\tau), z(\tau)) = 0$. In the last case, although $dx/d\tau$ is not necessarily zero, the trajectory on the ellipse $\cal E$ makes a bounce because the point $M={(x,y,z)}$ reaches the surface $\partial {\cal B}$ of the Berlingot.
Using the parametrization (\ref{eq.paramEll}), the Gram determinant $V^2(\tau)$ becomes a polynomial of degree 3 in $\cos\tau$ and $\sin\tau$. The zeros of $V^2(\tau)$ can thus be deduced from the roots of a polynomial of degree 6 in $\cos\tau$. Given that for non-stationary general solutions, the volume $V(\tau)$ (and the Gram determinant $V^2(\tau)$) only cancels at two different values $\tau_1$ and $\tau_2$, once the roots of $V^2(\tau)$ are known, one has to select $\tau_1$ and $\tau_2$ as the closest ones surrounding the initial condition $\tau_0={\rm atan}_2(z_0/A_z, x_0/A_x)$, where $x_0=x(t=0)$ and $z_0=z(t=0)$. Once the roots $(\tau_k)_{k=1,2}$ are known, the amplitudes of variation of $\theta_x$, $\theta_y$, and $\theta_z$ are straightforward. The angle $\theta_x$ varies between $\theta_{x,1} = \cos^{-1}x(\tau_1)$ and $ \theta_{x,2} = \cos^{-1} x(\tau_2)$, idem for $\theta_y$, and $\theta_z$. 
Nevertheless, the amplitude of $\theta_x$, $\theta_y$, or $\theta_z$ can be larger than what has just been calculated if there exists a value of $\tau$ between $\tau_1$ and $\tau_2$ such that $dx/d\tau=0$, $dy/d\tau=0$, or $dz/d\tau=0$, respectively. The derivative $dx/d\tau$ cancels at
\be
\tau_{x}^\pm = \frac{\pi}{2} \pm \frac{\pi}{2}\ ,
\ee
the quantity $dz/d\tau$ vanishes at
\be
\tau_z^{\pm} = 0 \pm \frac{\pi}{2}\ ,
\ee
and finally, $dy/d\tau=0$ for
\be
\tau_y^\pm = {\rm atan}_2(L A_x, -G' A_z) \pm \frac{\pi}{2}\ .
\ee
All values $\theta_x$ associated to $\tau_1$, $\tau_2$, and $\tau_x^\pm$ are extrema. Thus, if more than two such values exist ($\tau_x^+$ or $\tau_x^-$ falls between $\tau_1$ and $\tau_2$), one has to sort them to get the maximal amplitude of $\theta_x$. Idem for $\theta_y$ and $\theta_z$. For example, in the systems studied in the following sections, the condition $dy/d\tau=0$ is met several times with $\theta_y(\tau_1)<\theta_y(\tau_2)<\theta_y(\tau_y^-)$. In these cases, the variations of $\theta_y$ are thus bounded by $\theta_y(\tau_1)$ and $\theta_y(\tau_y^-)$.
\subsection{Nutation period}
\label{sec.periods}
The equations of motion (\ref{eq.evolswn}) expressed in terms of the dot products $x$, $y$, $z$ (\ref{eq.xyz}) are
\be
\EQM{
\dot{x} &= \nu_3 V z \ ; \crm
\dot{y} &= \nu_1 V x - \nu_4 V z \ ; \crm
\dot{z} &= -\nu_2 V x\ ,
}
\label{eq.evolxyz}
\ee
where $(\nu_k)_{k=1,\ldots,4}$ are frequencies defined in (\ref{eq.nu1234}).
It is then straightforward to check that the change of time leading to the parametrization (\ref{eq.paramEll}) satisfies
\be
d\tau = \sqrt{\nu_2\nu_3}Vdt\ .
\label{eq.taut}
\ee
The nutation period $P_{\rm nut}$ corresponds to the oscillation period of the dot products $x(t)$, $y(t)$, and $z(t)$, which is twice the time needed to reach $\tau_2$ starting from $\tau_1$, the two roots of $V^2(\tau)=0$ \citepalias{Boue_Laskar_Icarus_2006}. Using the definition of $\tau$ (\ref{eq.taut}), the result is
\be
P_{\rm nut} = \frac{2}{\sqrt{\nu_2\nu_3}} \int_{\tau_1}^{\tau_2}
\frac{d\tau}{V(\tau)}\ ,
\label{eq.Pn1}
\ee
where $V(x,y,z)$ is given by (\ref{eq.Vxyz}) and the para\-me\-tri\-za\-tion $x(\tau)$, $y(\tau)$, and $z(\tau)$ written in (\ref{eq.paramEll}). In the vicinity of the integral boundaries $(\tau_k)_{k=1,2}$, $V(\tau)$ evolves like $\sqrt{\abs{\tau-\tau_k}}$, unless the ellipse ${\cal E}$ is tangent to the Berlingot, a case we discard for the moment.  The integral (\ref{eq.Pn1}) is thus convergent \citepalias{Boue_Laskar_Icarus_2006}. Nevertheless, to avoid the singularities at the boundaries, one shall perform the following change of variable
\be
\tau = \frac{\tau_2-\tau_1}{2}\sin u+\frac{\tau_2+\tau_1}{2}
\ee
leading to
\be
P_{\rm nut} = \frac{\tau_2-\tau_1}{\sqrt{\nu_2\nu_3}}
\int_{-\pi/2}^{\pi/2} \frac{\cos u\,du}{V(u)}\ .
\label{eq.Pn}
\ee
If the trajectory of $(x,y,z)$ is tangent to the Berlingot `from the inside', the point of tangency is an unstable fixed point.  In that case, the nutation period $P_{\rm nut}$ becomes infinite as expected for an evolution to or from an equilibrium. On the other hand, if the trajectory is tangent `from the outside', the system is in the middle of a libration zone. In that case, the nutation period is derived from a linearization of the equations of motion (\ref{eq.evolswn}).
\subsection{Precession period}
As stated before, the precession period is more difficult to derive. We have reached the limit of the geometrical approach. It is now necessary to return to a more conventional description based on action/angle variables that we denote $(L_z, G_z, G'_z, \psi, \Omega, \Omega')$. The three actions are the third components of the angular momenta in a given reference frame while $\psi$, $\Omega$, and $\Omega'$ are the precession angle of the star and the longitudes of the ascending nodes of the planet system and of the companion, respectively. $(L_z, \psi)$ are Andoyer conjugate variables; $(G_z, \Omega)$ and $(G'_z, \Omega')$ are subsets of Delaunay variables. The system has thus {\em a priori} three degrees of freedom and the Hamiltonian written in these variables reads
\be
\EQM{
H =& -\frac{\alpha}{2}\left(\sin\varepsilon\sin I \cos(\psi-\Omega)
+\cos\varepsilon\cos I\right)^2 \crm &
     -\frac{\gamma}{2}\left(\sin I\sin I' \cos(\Omega-\Omega')
+\cos I\cos I'\right)^2\ ,
}
\ee
with $\cos\varepsilon = L_z/L$, $\cos I=G_z/G$, and $\cos I'=G'_z/G'$. We recall that $(L, G, G')$ are constants of motion. After a usual reduction of the node \citep{Malige_etal_CeMDA_2002}, the problem reduces to two degrees of freedom. The associated symplectic transformation is
\be
\left\{
\EQM{
\tilde \psi   &= \psi\ , \crm
\tilde \Omega &= \Omega - \psi\ ,\crm
\tilde \Omega' &= \Omega' - \psi\ ,
}\right.
\quad
\left\{
\EQM{
\tilde L_z &= L_z + G_z + G'_z\ ,\crm
\tilde G_z &= G_z\ ,\crm
\tilde G'_z &= G'_z\ ,
}\right.
\label{eq.canon}
\ee
and the Hamiltonian expressed in the new variables reads
\be
\EQM{
\tilde H =&
  -\frac{\alpha}{2}\left(\sin\varepsilon\sin I \cos\tilde \Omega
  +\cos\varepsilon\cos I\right)^2 \crm &
  -\frac{\gamma}{2}\left(\sin I\sin I' \cos(\tilde \Omega- \tilde \Omega')
  +\cos I\cos I'\right)^2\ ,
}
\label{eq.Hamtilde}
\ee
with $\cos\varepsilon = (\tilde L_z - \tilde G_z - \tilde G'_z)/L$, $\cos I= \tilde G_z/G$, and $\cos I'=\tilde G'_z/G'$. The variable $\tilde \psi$ is cyclic, its conjugated momentum $\tilde L_z$ is constant. $\tilde L_z$ represents the third component of the total angular momentum of the system. The two remaining degrees of freedom are carried by $(\tilde G_z, \tilde \Omega)$ and $(\tilde G'_z, \tilde \Omega')$.  The problem is further simplified when the reference plane is chosen to be the invariant plane orthogonal to the total angular momentum $\vec W$. Indeed, in that case the first two components of $\vec W$ are zero and thus the two angle variables $\tilde \Omega$ and $\tilde \Omega'$ are related by
\be
L\sin\varepsilon + G\sin I\exp(\ii\tilde \Omega) 
+G'\sin I'\exp(\ii\tilde \Omega') = 0\ .
\label{eq.third_integ}
\ee
Hence, the relative motion described by the Hamiltonian (\ref{eq.Hamtilde}) has only one degree of freedom when expressed in the invariant plane. This agrees with the fact that the relative motion is characterized by a single frequency called the nutation frequency. But although the Hamiltonian (\ref{eq.Hamtilde}) is integrable, the general solution does not seem easy to derive in action/angle variables. Next, we thus rely on the solution obtained in the variables $(x, y, z)$.
The expression of the precession period $P_{\rm prec}$ is then obtained by solving the evolution of $\tilde \psi = \psi$. Indeed, consider a 3-vector system in a given configuration at time $t=0$. After a nutation period $P_{\rm nut}$, the system returns to its initial configuration (same mutual distances) but its overall orientation is changed by an angle $\psi(P_{\rm nut})-\psi(0)$ around the total angular momentum. Hence,
\be
P_{\rm prec} = \frac{2\pi}{\Delta\psi} P_{\rm nut}\ ,
\ee
with $\Delta\psi = \abs{\psi(P_{\rm nut})-\psi(0)}$ is deduced from the equation of motion
\be
\frac{d\psi}{dt} = \frac{d\tilde \psi}{dt} = \Dron{\tilde H}{\tilde L_z} = \frac{1}{L} \Dron{\tilde H}{\cos\epsilon}\ .
\ee
Substituting the expression of the Hamiltonian (\ref{eq.Hamtilde}), and remembering that its first parenthesis (in factor of $\alpha/2$) is precisely $x=\vec s\cdot\vec w$, we get
\be
\frac{d\psi}{dt} = -\frac{\alpha}{L} x \left(-\frac{\cos\epsilon}{\sin\epsilon}\sin I\cos\tilde\Omega + \cos I\right)\ .
\ee
This expression is further simplified after expressing $\cos\tilde\Omega$ in terms of $x$, $\epsilon$, and $I$ as
\be
\frac{d\psi}{dt} = -\frac{\alpha}{L}x \frac{\cos I - x\cos \epsilon}{1-\cos^2\epsilon}\ .
\ee
Moreover, $\cos \epsilon$ and $\cos I$ are functions of $(x, y, z)$ given by
\be
\EQM{
\cos \epsilon & = \frac{\vec W\cdot \vec s}{W} = \frac{L + Gx + G'y}{W}\ , \crm
\cos I & = \frac{\vec W\cdot \vec w}{W} = \frac{Lx + G + G'z}{W}\ ,
}
\ee
where $W = \norm{\vec W}$ is the norm of the total angular momentum of the system and is a constant of the motion. Thus,
\be
\frac{d\psi}{dt} = -\frac{\alpha}{L}Wx\frac{G(1-x^2)+G'(z-xy)}{W^2-(L+Gx+G'y)^2}\ .
\label{eq.dpsidt}
\ee
Finally, using the change of time $t \mapsto \tau$ (\ref{eq.taut}) and the fact that the oriented volume $V(\tau)$ changes its sign at a rebounce, we get
\be
\EQM{
\Delta\psi &= \frac{1}{\sqrt{\nu_2\nu_3}}\abs{ \int_{\tau_1}^{\tau_2} \frac{d\psi}{dt}\,\frac{d\tau}{V(\tau)} - \int_{\tau_2}^{\tau_1} \frac{d\psi}{dt}\,\frac{d\tau}{V(\tau)}}\ , \crm
&= \frac{2}{\sqrt{\nu_2\nu_3}}\abs{ \int_{\tau_1}^{\tau_2} \frac{d\psi}{dt}\,\frac{d\tau}{V(\tau)}}\ .
}
\ee
The expression of the precession period is thus
\be
P_{\rm prec} = \pi \sqrt{\nu_2\nu_3} P_{\rm nut} 
\abs{ \int_{\tau_1}^{\tau_2} \frac{d\psi}{dt}\,\frac{d\tau}{V(\tau)}}^{-1} \ ,
\label{eq.Pp}
\ee
with $d\psi/dt$ given by (\ref{eq.dpsidt}).

\section{$N$-body problem}
\label{sec.nbodyproblem}
\subsection{Numerical model}
The generic system we wish to study is composed of $p$ packed planets orbiting an oblate star with a companion on a wide eccentric inclined orbit. The planet system is supposed to be relatively coplanar with low eccentricities. Furthermore, we assume that mean-motion resonances are not dominating the secular motion. In that case, using the formalism developed in (BF14), the secular Hamiltonian of the system can be approximated by
\be
\EQM{
H_{\rm tot} =& \sum_{1\leq j<k\leq p} \bar{H}_{\rm close}(j,k) + \sum_{j=1}^p \Big(
\bar H_{\rm hierar}(j, p+1)
\crm &+ \bar H_{\rm spin}(0, j) + \bar H_\mathrm{relat}(j) \Big)\ .
}
\label{eq.Htot}
\ee
In this equation, $\bar H_\mathrm{close}(j,k)$ represents the secular interaction between planets $j$ and $k$ expanded up to the fourth order in eccentricity and mutual inclination. $\bar H_{\rm hierar}(j,p+1)$ denotes the secular interaction between a planet $j$ and the companion. This term is developed in semimajor axis ratio at the octupole. $\bar H_{\rm spin}(0, j)$ is the quadrupolar secular spin-orbit interaction between the stellar spin-axis and the planet $j$. Finally, $\bar H_{\rm relat}(j)$ describes the relativistic precession of the pericenter of the planet $j$ induced by the central star. The expression of each of these terms as well as the corresponding equations of motion are detailed in (BF14). Hereafter, we refer to numerical integrations of $H_{\rm tot}$ as the {\em numerical} model. Note that because the Hamiltonian $H_{\rm tot}$ is averaged over all mean anomalies, the $N$-body problem (with $N=p+2$) has already been changed into an $N$-ring problem (including the equator of the central star).
In the following, we show how the numerical $N$-ring problem can be further simplified into a $3$-vector problem which will be referred to as the {\em analytical} model. Then, we discuss the validity of the latter and perform a few comparisons of the two models. 

\subsection{Analytical model}
The 3-vector problem presented at the beginning provides the solution of the secular evolution of obliquity and inclinations only. Conversely, the numerical model also accounts for the evolution of eccentricities. The analytical model is thus more limited. Actually, it uses the fact that eccentricities and inclinations are decoupled in Laplace-Lagrange linear theory, and that the quadrupole expansion of hierarchical interaction is independent of the direction of the outer body's pericenter, which is known as the {\em happy coincidence} \citep{Lidov_Ziglin_CeMDA_1976}. At this order, and neglecting planet eccentricities, the Hamiltonian $H_{\rm tot}$ given in (BF14) reduces to
\be
\EQM{
H_{\rm inc}^{(2)} =& - \sum_{j=1}^p \frac{\alpha_j}{2}(\vec s\cdot \vec w_j)^2
-\sum_{j=1}^p \frac{\gamma_j}{2}(\vec w_j\cdot\vec w')^2
\crm &
-\sum_{1\leq j<k\leq p} \beta_{jk}(\vec w_j\cdot\vec w_k)\ ,
}
\label{eq.Hinc}
\ee
where $\vec w_j$ is the unit vector normal to the orbit $j$ and
\be
\EQM{
\alpha_j &= \frac{3}{2}\frac{{\cal G}m_0m_jJ_2R_0^2}{a_j^3}\ , \crm
\gamma_j &= \frac{3}{4}\frac{{\cal G}m_jm'a_j^2}{a'^3(1-e'^2)^{3/2}}\ , \crm
\beta_{jk} &= \frac{1}{4}\frac{{\cal G}m_jm_ka_j}{a_k^2}\lap{3/2}{1}\left(\frac{a_j}{a_k}\right)\ .
}
\label{eq.alpbetgam}
\ee
Hereafter, $m$ denotes masses, $a$ semimajor axes, and $e$ eccentricities. Subscripts $0$ and $j=1,\ldots,p$ stand for the central star and for the $p$ planets, respectively. The companion's parameters are written with a prime. ${\cal G}$ is the gravitational constant. The star has a radius $R_0$, a second fluid Love number $k_2$, a rotation speed $\omega_0$, a moment of inertia $C$, and a quadrupole coefficient $J_2$ given by \citep[e.g.,][]{Lambeck_book_1988}
\be
J_2 = k_2\frac{\omega_0^2 R_0^3}{3{\cal G}m_0}\ .
\label{eq.Jom}
\ee
$\lap{s}{k}$ are Laplace coefficients. Note that all notations are summarized in Tab.~\ref{tab.notation}.
Let $\vec L=L\vec s=C\omega_0\vec s$, $\vec G_j=G_j\vec w_j$, and $\vec G'=G'\vec w'$ be the angular momenta of the star, of the orbit of planet $j$, and of the orbit of the companion, respectively. Within the same level of approximation as in Eq.~(\ref{eq.Hinc}), $e'$ and thus $G'$ are constant and the equations of motion are
\be
\EQM{
\frac{d\vec s}{dt}   &= -\frac{1}{L} \vec s \times \grad{\vec s}H_{\rm inc}^{(2)}\ , \crm
\frac{d\vec w_j}{dt} &= -\frac{1}{G_j} \vec w_j \times \grad{\vec w_j}H_{\rm inc}^{(2)}\ , \crm
\frac{d\vec w'}{dt}  &= -\frac{1}{G'} \vec w' \times \grad{\vec w'}H_{\rm inc}^{(2)}\ .
}
\label{eq.swjw'}
\ee
In a last step, we consider the total angular momentum $\vec G$ of the planet system
\be
\vec G = \sum_{j=1}^p G_j \vec w_j\ .
\label{eq.vecG}
\ee
We also denote $G = \norm{\vec G}$ and $\vec w = \vec G/G$. The latter represents the normal of the planet system. Using the equations of motion (\ref{eq.swjw'}), we get
\be
\frac{d\vec G}{dt} = - \sum_{j=1}^p \big(\alpha_j (\vec s\cdot \vec w_j)\vec s 
                                   + \gamma_j (\vec w'\cdot\vec w_j)\vec w'\big)\times \vec w_j\ .
\label{eq.dvecGdt}
\ee
As expected, all the interactions internal to the planet system (involving $\beta_{jk}$ and $\beta_{kj}$) cancel out in the evolution of $\vec G$. Furthermore, assuming that each planet $j$ remains close to the planet system (low mutual inclination), $\vec w_j$ is replaced by $\vec w$ in (\ref{eq.swjw'}) and (\ref{eq.dvecGdt}). As a result, $G$ becomes constant and the equations of motion read
\be
\EQM{
\frac{d\vec s}{dt} &= - \frac{\alpha}{L} (\vec s\cdot \vec w) \vec w\times \vec s\ , \crm
\frac{d\vec w}{dt} &= - \frac{\alpha}{G} (\vec s\cdot \vec w) \vec s\times \vec w 
                      - \frac{\gamma}{G} (\vec w'\cdot\vec w) \vec w'\times\vec w\ , \crm
\frac{d\vec w'}{dt}&= - \frac{\gamma}{G'}(\vec w'\cdot\vec w) \vec w\times\vec w'\ ,
}
\label{eq.dsww}
\ee
with $\alpha = \sum_j \alpha_j$ and $\gamma = \sum_j \gamma_j$. These equations are exactly those of the three vector problem (\ref{eq.evolswn}). We can thus use the results of the 3-vector problem to study the spin-orbit evolution in compact planetary systems perturbed by an inclined companion. The expressions of the characteristic frequencies $\nu_k$, Eq.~(\ref{eq.nu1234}), are
\be
\EQM{
\nu_1 &=& \frac{\pi k_2}{P_0} \frac{m_0 R_0^2}{C} A_1 \ ,\crm
\nu_2 &=& \frac{\pi k_2}{P_0} \frac{A_1}{A_2}\ ,\crm
\nu_3 &=& \frac{3\pi}{2P'} \frac{m'}{m_0+m'} \frac{P_0}{P'} 
   \frac{A_3}{A_2}
   \frac{1}{(1-e'^2)^{3/2}} \ ,\crm
\nu_4 &=& \frac{3\pi}{2P'} \left(\frac{R_0}{a'}\right)^2
   \frac{A_3}{(1-e'^2)^2} \ ,
}
\label{eq.nus}
\ee
with
\be
\EQM{
A_1 &= \sum_{j=1}^P \frac{m_j}{m_0}\left(\frac{R_0}{a_j}\right)^3\ ,\crm
A_2 &= \sum_{j=1}^p \frac{m_j}{m_0+m_j}\frac{P_0}{P_j} \left(\frac{a_j}{R_0}\right)^2\ ,\crm
A_3 &= \sum_{j=1}^p \frac{m_j}{m_0}\left(\frac{a_j}{R_0}\right)^2\ .
}
\label{eq.An}
\ee
In (\ref{eq.nus}) and (\ref{eq.An}), $P_0$, $P_j$, and $P'$ are the rotation period of the central star, the orbital period of the planet $j$, and the orbital period of the companion, respectively.
It must be stressed that although the model (\ref{eq.dsww}) has been obtained for a non-resonant planet system, the result is more general. Indeed, the equation (\ref{eq.dvecGdt}) holds whatever are the interactions between the planets: the evolution of $\vec G$ depends solely on external perturbations (those raised by the stellar oblateness and by the companion). As a consequence, the 3-vector approximation also applies to problems with a resonant planetary system as long as their eccentricities and mutual inclinations remain low. But the same is not true for the numerical model because these planetary orbital elements are explicitly integrated using non-resonant secular equations.
\subsection{Validity of the analytical model}
Here, we assume that the planet system is non-resonant. Then, both the numerical and the analytical models are valid as long as the eccentricities and the mutual inclinations of the planets remain small. This condition is easy to check with the numerical one because it includes the evolution of the planet eccentricities and their coupling with inclinations. Conversely, the analytical model completely discard planetary eccentricities and individual inclinations. As a result, the latter does not provide any information on its validity. It would thus be convenient to have a simple criterion saying whether or not eccentricities might evolve significantly. Stated in different words, the criterion should distinguish among all systems those which are dynamically rigid.
In the case with only two planets, the Lidov-Kozai cycles are suppressed as long as the periapsis precession period of the outer planet $\tau_{\rm pp}$ is shorter than the Lidov-Kozai timescale $\tau_{\rm Koz}$ \citep{Takeda_etal_ApJ_2008}. The study of the general case is beyond the scope of this paper. We thus assume that this is also true in the $p$-planet case. 
Under the approximation of the Laplace-Lagrange theory, the secular frequencies of multiplanet systems are those of the eigenmodes. However, in order to get a simple criterion and to avoid the problem of assigning an eigenfrequency to a specific planet, we only consider the secular interaction between the two outermost planets. More precisely, within the Laplace-Lagrange linear model, the secular evolution of $\vec e_p$ induced by the planet $p-1$, with the notation of (BF14), reads
\be
\frac{d\vec e_p}{dt} =  \frac{{\cal G}m_{p-1}m_p}{a_pG_p}
                        \Big(c_2\, \vec w\times \vec e_p 
                          +  c_3\, \vec w\times \vec e_{p-1} \Big)
\label{eq.depdt}
\ee
where $G_p = m_0m_p/(m_0+m_p)\sqrt{{\cal G}(m_0+m_p)a_p}$ and $c_2$ and $c_3$ are two functions of $(a_{p-1}/a_p)$. We then set $\tau_{\rm pp}$ as the inverse of the coefficient of $\vec w\times\vec e_p$ in (\ref{eq.depdt}). At the lowest order in $(a_{p-1}/a_p)$, the result is
\be
\frac{\tau_{\rm pp}}{P_p} \approx \frac{2}{3\pi}\frac{m_0}{m_{p-1}}\left(\frac{a_p}{a_{p-1}}\right)^{2}\ ,
\label{eq.taupp}
\ee
with $P_p$ the orbital period of the (outermost) planet $p$. On the other hand, the Lidov-Kozai timescale \citep{Lidov_Ziglin_CeMDA_1976}, written in canonical astrocentric elements is
\be
\frac{\tau_{\rm Koz}}{P_p} = \frac{2}{3\pi}\frac{m_0}{m'}\left(\frac{b'}{a_p}\right)^3\ ,
\label{eq.tauKoz}
\ee
where $b'=a'\sqrt{1-e'^2}$ is the semiminor axis of the companion.  The criterion of validity of the analytical model is thus
\be
\tau_{\rm pp} \ll \tau_{\rm Koz}\ ,
\ee
where $\tau_{\rm pp}$ and $\tau_{\rm Koz}$ are given by (\ref{eq.taupp}) and (\ref{eq.tauKoz}), respectively.
\figValidity
We have checked this criterion on 4-body problems through 4\,800 integrations of the numerical model. The simulated system is composed of two planets with mass $10^{-5}M_\odot$ orbiting a $1M_\odot$ star. We chose three different masses for the companion: $m'=10^{-3}$, $10^{-2}$, $10^{-1}M_\odot$, two eccentricities $e'=0$, $0.5$, and two inclinations relative to the initial planet plane $I'_0 = 50$, $89^\circ$. For each combination $(m', e', I'_0)$, the semimajor axis of the inner planet and the semiminor axis of the companion take 20 values uniformly distributed in the ranges $0.1\leq a_{p-1}/a_p\leq 0.9$ and $10\leq b'/a_p \leq 100$, respectively. Neither spin-orbit interaction nor general relativity are included in this analysis. For $I'_0=89^\circ$, we found a sharp transition between low ($\max(e_p)<0.01$) and high ($\max(e_p)>0.8$) eccentricity regimes at $\tau_{\rm Koz} \sim 1.5 \tau_{\rm pp}$ leading to the criterion
\be
\frac{b'}{a_p} \gtrsim 1.5 \left(\frac{m'}{m_{p-1}}\right)^{1/3}\left(\frac{a_{p-1}}{a_p}\right)^{-2/3}
\label{eq.validity}
\ee
represented in Fig.~\ref{fig.validity}. For $I'_0=50^\circ$, the amplitude of $\max(e_p)$ (not shown) is smaller than in the $I'_0=89^\circ$ case. Nevertheless, a sharp transition is still observed although at lower values of $b'$ (the limits obtained with the two inclinations differ by 25\%). The constraint given by Eq.~(\ref{eq.validity}) is thus more stringent.
Finally, we would like to stress that the criterion (\ref{eq.validity}) displayed in Fig.~\ref{fig.validity} has only been verified in limited configurations. For instance, we have only considered inner systems composed of two planets with equal mass. We expect the criterion to remain informative in other cases, but close to its threshold we recommend integrations of the numerical model to certify the validity of the analytical one. Moreover, the criterion has to be adapted if the system contains several well separated compact groups of planets. It is indeed necessary to check whether they all behave coherently or not.

\section{Numerical tests and applications}
\label{sec.numerical}
In the previous paper BF14, we tested the numerical model against N-body integrations, finding them to hold well in for a system like the solar system, with low eccentricities and inclinations, and hierarchical system  which is highly inclined and undergoing Kozai-Lidov oscillations.  In this section, we assume the numerical model can well represent the motion of planetary systems, and we use it to test the analytic model in a few cases involving two observed exoplanet systems.  
The first one, 55~Cancri, is composed of five planets orbiting one member of a wide binary system.  This system is actually the prototype for which the formalism of the previous section has been developed. The second system, HD~20794, contains three super-Earths, but no wide perturber detected to date.  We have thus added an arbitrary planetary companion to it.  This latter system has been chosen because of the low angular momentum in the planets orbit in comparison to the 55~Cancri system. The two systems allow to explore relatively different regions of the parameter 
space.
\subsection{55 Cancri}
\label{sec.55Cnc}
The system 55~Cnc is our first example because it is actually a binary system and five planets orbit the most massive component 55~Cnc~A.  Moreover, the evolution of the inclination of the planet system has already been studied in \citet{Kaib_etal_ApJ_2011} (hereafter noted \citetalias{Kaib_etal_ApJ_2011}) using another approach. The masses and the orbital parameters of the five planets are summarized in Tab.~\ref{tab.Cnc}. Planets b and c are close to the 3:1 mean-motion resonance \citep{Fischer_etal_ApJ_2008}, but we assume that this proximity does not affect significantly the secular motion derived from the numerical model. As in \citetalias{Kaib_etal_ApJ_2011}, we introduce small initial inclinations to avoid purely coplanar evolutions of the planet system.
\TabCnc
The central star 55~Cnc~A has a mass $m_0=0.905M_\odot$, and a radius $R_0=0.943R_\odot$ \citep{vonBraun_etal_ApJ_2011}. Its rotation period determined photometrically is $P_0=42.7$ days \citep{Fischer_etal_ApJ_2008}.  The internal structure parameters, $k_2 = 0.028$ and $C=0.09\,m_0R_0^2$, have been taken from stellar models \citep[][Tab.~1a]{Landin_etal_AA_2009}.  The stellar companion is at a projected separation of 1065 au and has a mass $m'=0.26M_\odot$ \citep{Mugrauer_etal_AN_2006}. In all the following studies, the initial obliquity of the central star relative to the planet plane $\theta_x(t=0)$ is set to zero.
As a first test, we reproduce the evolution presented in \citetalias{Kaib_etal_ApJ_2011}'s Fig.~1.  The orbital parameters of the companion are $a'=1250$ au, $e'=0.93$, and $I'=115$ deg. For this simulation, we use the numerical Hamiltonian expressed in vectorial form (BF14).  Thus planet-planet interactions are modeled at the fourth order in eccentricity inclination, the interaction with the companion is modeled at the third order in semi-major axis ratio. Relativistic precession is included, but the central star is considered as a point mass, we thus neglect the interactions with the stellar figure. The result of the simulation is displayed in Fig.~\ref{fig.Kaiba}. 
\figKaiba
The numerical evolution agrees very well with the $n$-body integration performed in \citetalias{Kaib_etal_ApJ_2011}. All the planet orbital inclinations follow the same track. The planetary orbital plane tilts periodically like a rigid body. The main difference between our work and \citetalias{Kaib_etal_ApJ_2011}'s result is a slight shift in the precession frequency. In \citetalias{Kaib_etal_ApJ_2011}'s figure, there are twelve full precession oscillations within 1 Gyr, while in ours the twelfth one is not finished.
Now, we focus our attention on the evolution of the spin axis of the central star. Figure~\ref{fig.Kaiba}, showing the solid rotation of the planetary system, justifies the analytical model of the previous section: the planetary system remain almost coplanar. The numerical values of the precession frequencies, Eq.~(\ref{eq.nus}), are $\nu_1 \approx 323$ deg/Myr, $\nu_2 \approx 0.49$ deg/Myr, $\nu_3 \approx 9.8$ deg/Myr, and $\nu_4 \approx 0.030$ deg/Myr. The equations of motion are thus
\be
\EQM{
\frac{d\vec s}{dt} &= -323 \cos\theta_x\ \vec w\times \vec s\ ; \crm
\frac{d\vec w}{dt} &= -0.49 \cos\theta_x\ \vec s\times \vec w 
                      -9.8  \cos\theta_z\ \vec w' \times \vec w\ ; \crm
\frac{d\vec w'}{dt} &= -0.030 \cos\theta_z\ \vec w \times \vec w'
}
\ee
in deg/Myr. The precession motion of the spin axis $\vec s$ of the central star relative to the pole $\vec w$ of the planet mid-plane is about eighty times faster than any of the other secular motions (when taking into account $\cos \theta_z\approx -0.42$). In good approximation, the evolution of $\vec s$ can thus be computed assuming all the other vectors as constant. As a result, the obliquity $\theta_x$ of the star with respect to the orbital plane of the planets should remain constant and equal to its initial value. If it begins aligned, it will remain aligned. A numerical integration similar to that of Fig.~\ref{fig.Kaiba}, but including the evolution of the orientation of the central star (represented by the vector $\vec s$), confirms this analytical conclusion.  The trajectory of the unit vectors $\vec s$, $\vec w$, and $\vec w'$ are plotted in Fig.~\ref{fig.Kaibb}{\bf a},{\bf c} as well as $\vec w_1$ to check whether the innermost planet of the system follows the evolution of the outer planets. These trajectories are a combination of a solid rotation (precession motion) around the total angular momentum and quasi-periodic nutation motions \citepalias{Boue_Laskar_Icarus_2006, Boue_Laskar_Icarus_2009}. In Fig.~\ref{fig.Kaibb}{\bf b},{\bf d}-{\bf f}, the trajectories, displayed in the frame rotating with the main precession frequency, show that the stellar spin axis $\vec s$ never moves away from $\vec w$ and $\vec w_1$ by more than about $2^\circ$. For comparison, the analytical model (section~\ref{sec.amplitude_nutation}) predicts a spin-orbit amplitude of $1.4^\circ$. The star is thus pulled along with the planetary orbits as their inclinations oscillate.  As a consequence, the sole presence of an inclined stellar companion in the 55 Cnc system is not enough to generate a spin-orbit misalignment.
\figKaibb
The above conclusions have been derived for a specific set of physical and orbital parameters. Is there any other choices producing effective misalignment ? In this system, the star follows the tilt of the orbits of the planets because $\nu_1 \gg \nu_3$. To cancel this effect, one should decrease $\nu_1$ or increase $\nu_3$. 
Except for the planet e observed in transit, only the minimum masses $m_k\sin i_k$, where the inclinations $i_k$ are measured with respect to the plane of the sky, are known. Nevertheless, the frequency $\nu_1$ depends linearly on $m_k$ while $\nu_3$ is not affected by a rescaling of the planet masses. Thus, increasing the planet masses would only cause the star to be more strongly coupled to the planets.
In the expression of $\nu_1$ (\ref{eq.nus}), the next less known parameter is the Love number $k_2$ of the star. Based on the results of the internal structure model of \citet{Landin_etal_AA_2009}, we estimate the error on $k_2$ to be of the order of 20\%. But even if $k_2$ is wrong by a factor two, it is not enough to compensate for the large ratio $\nu_1/\nu_3\sim 33$. 
On the contrary, the precession frequency $\nu_3$ is much more undetermined. Indeed, $\nu_3$ scales as $m' / [ a'^{3} (1-e'^2)^{{3/2}}]$. The mass $m'=0.26M_\odot$ has been derived by \citet{Mugrauer_etal_AN_2006} from the absolute infrared magnitude using the evolutionary models of \citet{Baraffe_etal_AA_1998}, whereas the semimajor axis $a'=1025$ au and the eccentricity $e'=0.93$ are only constrained by the projected separation $d=1065$ au. There is thus a full range of semimajor axes and eccentricities compatible with the observations. The closer and the more eccentric the perturber's orbit is, the larger is the frequency $\nu_3$. Fixing the distance of the pericenter at $q=35$ au, and assuming that the projected separation corresponds to the distance of the apocenter $Q=1065$ au, we get $a'=550$ au, and $e'=0.936$. Moreover, since $\nu_3$ is proportional to $\cos\theta_z$, where $\theta_z$ is the orbital inclination of the stellar companion relative to the planetary mid-plane, we chose a lower initial inclination $I'(t=0) = \theta_z(t=0) = 30^\circ$ only (we recall that $I'$ is measured with respect to the reference plane, which is also the planet plane at $t=0$, while $\theta_z$ is the mutual inclination between the orbits of the companion and of the planets). With this set of parameters, $\abs{\nu_3\cos\theta_z}$ increases from 4.2 deg/Myr to 114 deg/Myr.  This quantity is still less that $\nu_1$ but it is of the same order of magnitude. The trajectories of the unit vectors computed with this new set of initial conditions are plotted in Fig.~\ref{fig.Kaibc}. We indeed observe that the stellar axis and the planets orbital pole gets periodically misaligned from $0^\circ$ to about $50^\circ$. In this case, the amplitude of $32^\circ$ given by the analytical model is underestimated by about 36\%. This discrepancy is mainly due to the relatively weak coupling between 55~Cnc~d and the four innermost planets. The semimajor axis ratio between the two outermost planets is about 7.4. There is a better agreement on the nutation period derived from the numerical integration 2.3~Myr, and analytically, 1.8~Myr.
\figKaibc
We have shown on this example (55~Cancri) that the central star can be strongly linked to the motion of the orbital plane of its planets. This is mainly due to the fact that most of the angular momentum is in the orbit of the planets rather than in the rotation of the star.  Nevertheless, if the interaction with the outer companion is strong enough, the evolution of the stellar spin axis can be decoupled from the motion of its planet. A detailed analysis of the statistical distribution of the spin-orbit misalignment in the 55~Cnc system is beyond of the scope of the present paper. Nevertheless, a global analysis of the maximal spin-orbit misalignment of compact systems perturbed by a wide companion is performed in section~\ref{sec.global}.
\subsection{HD~20794}
The system HD~20794 is composed of three super-Earths orbiting a G8V type star with mass $m_0=0.85M_\odot$ and radius $R_0=0.9R_\odot$ \citep{Pepe_etal_AA_2011,Bernkopf_etal_MNRAS_2012}.  In comparison to 55 Cancri, this system is not known to harbor any outer companion. We nevertheless choose this system because the angular momentum is more balanced between the star and the planets than in the 55 Cancri system, and this difference allows new kinds of spin-orbit angle evolutions.  The rotation period of the star is $P_0=33.2$ days \citep{Pepe_etal_AA_2011}, and its internal structure constants inferred from \citet{Landin_etal_AA_2009} are $k_2=0.017$ and $C=0.08 m_0 R_0^2$.  The parameters of the three super-Earths are summarized in Tab.~\ref{tab.HD}.
\TabHD
Because no stellar companion has been detected in this system until now, we arbitrarily add a perturber: a giant planet with mass $m'=1M_J$ at $a'=20$ au with an eccentricity $e'=0.1$ on an orbit initially inclined by $30^\circ$ with respect to the plane of the planets.  This planet would have a 97 yr orbital period and a radial velocity semi-amplitude $K=7.25$ m s$^{-1}$. For an observation time span of 7 years \citep{Pepe_etal_AA_2011}, the amplitude of the drift induced by the new planet would range between 0.19 m s$^{-1}$ and 3.3 m s$^{-1}$, depending on the orbital phase.  Although the upper limit is well above the 0.82 m s$^{-1}$ rms of the residuals in \citet{Pepe_etal_AA_2011}, the lower limit is significantly below and such a planet could have been missed. In this system, the central star and the planet system have very similar angular momenta.  Thus, the frequencies $\nu_1$ and $\nu_2$ are almost equal. The equations of motion are
\be
\EQM{
\frac{d\vec s}{dt} &= -0.211 \ \vec w \times \vec s\ ; \crm
\frac{d\vec w}{dt} &= -0.164 \ \vec s \times \vec w 
                      -6.12 \ \vec w' \times \vec w\ ; \crm
\frac{d\vec w'}{dt} &= -0.025 \ \vec w \times \vec w'
}
\label{eq.hdinit}
\ee
in deg/Myr. From (\ref{eq.hdinit}), we deduce that the outer giant planet possesses most of the angular momentum of the system, its orbit remains close to the invariant plane of the system. The equations (\ref{eq.hdinit}) also tell us that the angular momentum of the planet system $\vec w$ precesses around $\vec w'$ about 26 times faster than the star $\vec s$ around $\vec w$ (with $\cos\theta_z\approx0.87$). As a consequence, the star precesses around the averaged orbital pole $\moy{}{\vec w}$ which is collinear to $\vec w'$, but at a much slower frequency than $\vec w$.  Figure~\ref{fig.HDa} shows the trajectories of $\vec s$, $\vec w_1$, $\vec w$, and $\vec w'$ in the frame rotating at the precession frequency of the stellar axis $\vec s$ around the total angular momentum.
\figHDa
Since the precession motion of the planets pole is faster than that of the star, the planets mid-plane is still precessing around the total angular momentum of the system in the rotating frame. As a consequence, the spin-orbit angle oscillates periodically between $0^\circ$ and $60^\circ$, the latter being equal to twice the initial inclination $I'(t=0)$. This amplitude is in perfect agreement with the analytical model, as well as the nutation period which is equal to 67.4~Myr.
\figHDb
In a last experiment, we reduce the mass $m'$ of the perturbing companion by a factor 25, such that the frequencies $\nu_3$ becomes of the same order of magnitude as $\nu_1$ and $\nu_2$. As a result, $\vec w$ is expected to describe a nutation motion around a center located half-way between $\vec s$ and $\vec w'$. Moreover, the amplitude of the nutation of $\vec s$ should be similar to that of $\vec w$. This is confirmed by the Figure~\ref{fig.HDb} showing the trajectories of the unit angular momenta in the frame rotating at the precession frequency. Because of the initial condition $\theta_x(t=0)=0^\circ$, the nutation loops of $\vec w$ and $\vec s$ are tangent. In that case, the spin-orbit angle $\theta_x$ between the stellar axis and the planets orbit pole oscillate between $0^\circ$ and $36^\circ$ at a much longer nutation period of 1.07~Gyr. The analytical model reproduces exactly these two quantities (amplitude and period).
\section{Global analysis}
\label{sec.global}
In the previous section, the application of the numerical and the analytical models on a few systems revealed different types of evolution. Here, we exploit more deeply the geometric structure of the analytical model to derive very general results. In particular, we abandon studies of individual motions in favor of more global analysis of group of trajectories. In all cases, we assume that the initial spin-orbit angle $\theta_x(t=0)$ is nil.
\subsection{Hyperboloid of trajectories}
\label{sec.hyperboloid}
In the analytical approximation, section~\ref{sec.integrable}, systems composed of packed planets surrounded by a perturber on a wide orbit have fixed physical and orbital parameters except inclinations and obli\-quities. The constant orbital parameters are the semimajor axes of the planets and the semiminor axis of the perturber $b'=a'\sqrt{1-e'^2}$.  It is thus natural to keep masses, semimajor axes and eccentricities at given values, and study the effect of initial inclinations on the behavior of those systems. Furthermore, considering the hypothesis where the initial stellar obliquity $\theta_x(t=0)$ is nil, i.e., $x=1$ and $y=z$ (${\cal D}_x$ in Fig.~\ref{fig.Berl}), evolutions are only characterized by the initial inclination of the perturber $I'_0=I'(0)=\theta_y(0)=\theta_z(0)$ with respect to the inner system. Thus, in the following we consider the surface resulting from the union of the trajectories in the $(x,y,z)$ frame with $0\leq I'_0 \leq 180^\circ$, or equivalently, $-1\leq z_0\leq1$.  More generally, we define ${\cal S}$ as the set of all $(x,y,z)\in\mathbb{R}^3$ such that there exists a $z_0\in\mathbb{R}$ verifying 
\be
\EQM{
H(x,y,z) &= H(1,z_0,z_0) \crm
K(x,y,z) &= K(1,z_0,z_0)
}
\label{eq.defS}
\ee
where $H$ (\ref{eq.Hxyz}) and $K$ (\ref{eq.Kxyz}) are the two first integrals of the motion. The second equation of (\ref{eq.defS}) provides the expression of $z_0$ as a function of $(x,y,z)$
\be
z_0 = p (x-1) + q y + r z\ ,
\label{eq.z0}
\ee
with
\be
\EQM{
p &= \frac{LG}{(L+G)G'}\ ,\crm
q &= \frac{L}{L+G}\ ,\crm
r &= \frac{G}{L+G}\ .
}
\ee
The substitution into the first equation of (\ref{eq.defS}) gives
\be
{\cal S} = \{(x,y,z)\in \mathbb{R}^3\mid S(x,y,z)=0\}\ ,
\ee
with
\be
S = \frac{\alpha}{\gamma} (x^2-1) + z^2-
\big(p (x-1) + q y + r z\big)^2\ .
\label{eq.Sxyz}
\ee
$S$ being a quadratic function of $(x,y,z)$, ${\cal S}$ is a quadric surface. Moreover, the diagonal ${\cal D}_x$ belongs to ${\cal S}$.  Therefore, ${\cal S}$ is an hyperboloid of one sheet (neither ellipsoid nor hyperboloid of two sheets contain a straight line). For some peculiar values of the parameters, ${\cal S}$ can also be a cylinder or a cone, but this set of parameters is negligible in the sense that its measure is zero. In the approximation $G'\gg (L,G)$ (or $p\ll (q,r)$), the case in which the outer body dominates the system's angular momentum, $S$ simplifies into
\be
S \approx \big(x\ y\ z\big)
[S]
\bpm x \cr y \cr z \epm - 1\ ,
\label{eq.approxS}
\ee
where
\be
[S] =
\bpm 1 & 0 & 0  \cr
     0 & -q^2\gamma/\alpha & -qr\gamma/\alpha \cr
     0 & -qr\gamma/\alpha & (1-r^2)\gamma/\alpha \epm\ .
\ee
The matrix $[S]$ has at least one positive eigenvalue (equal to 1), and its determinant $-(\gamma q/\alpha)^2$ is negative.  There is thus exactly two positive and one negative eigenvalues, and ${\cal S}$ is indeed an hyperboloid of one sheet.  Such an hyperboloid is plotted in Fig.~\ref{fig.quadric} with the parameters of the system 55~Cancri and where $b'=190$ au. The figure also shows the cube ${\cal C} = [-1,1]^3$ as well as a section of an elliptic trajectory in red whose initial conditions are $x(0)=1$ and $y(0)=z(0)=\cos^{-1}45^\circ$ meaning that the companion is initially inclined by 45$^\circ$ with respect to the equator of the star and the planet system mean plane. 
\figQuadric
As expected, the diagonal ${\cal D}_x$ representing the locus of all the initial conditions (see Fig.~\ref{fig.Berl}) belongs to the hyperboloid ${\cal S}$ (\ref{eq.Sxyz}), and is represented by the thick black line in Fig.~\ref{fig.quadric}. Within the approximation (\ref{eq.approxS}), this hyperboloid is centered on the origin, and given that $r=1-q$ with $0\leq q\leq 1$, it can be shown that the axes of symmetry are deduced from $(Ox, Oy, Oz)$ by a rotation of an angle $0\leq \chi \leq \pi/8$ around the $x$-axis.  The limit $\chi=\pi/8$ is reached when $q\rightarrow 0$, i.e., when planets have more angular momentum than the star ($G\gg L$), which is the case in Fig.~\ref{fig.quadric}, while the limit $\chi=0$ corresponds to $q\rightarrow 1$ where most of the angular momentum is in the star ($L\gg G$). We stress that the approximate equation (\ref{eq.approxS}) defining ${\cal S}$ has been provided in order to get an idea of the shape and of the orientation of the hyperboloid.  In Fig.~\ref{fig.quadric} and in the following, we always use the exact expression (\ref{eq.Sxyz}).
\subsection{Intersection with the Berlingot and maximal obliquity}
\figSB
Now that the quadric ${\cal S}$ is determined, we examine its intersections with the surface of the Berlingot $\partial {\cal B}$. The maximal amplitudes of the spin-orbit angle $\theta_x$ are then deduced from these intersections. To make the approach more concrete, we use again the two examples of section~\ref{sec.numerical}.
The two surfaces $\cal S$ and $\partial \cal B$ are plotted in figure \ref{fig.SB} with parameters of 55~Cancri for $b'$=190, 182, 180, and 170 au. As in Fig.~\ref{fig.quadric}, the initial conditions are the thick black diagonal ${\cal D}_x$. In the case $b'=190$ au (panel {\bf a}), and for prograde orbits like the red curve, elliptic trajectories evolve through the Berlingot, toward decreasing values of $x$ and $y$ (increasing spin-orbit angles between the star and both the planet system and the companion). Trajectories exit the Berlingot after a relatively short distance and reenter the Berlingot at negative values of $x$ and $y$ (retrograde rotation of the star) before exiting again.  Because the evolution of the system is restricted to the interior of the Berlingot, the maximal spin-orbit angles correspond to the first exit from the Berlingot. This topology is equivalent to the case studied by \citet{Kaib_etal_ApJ_2011}. As the distance of the companion decreases (panel {\bf b}, {\bf c}, and {\bf d}), the hyperboloid shrinks so that the first exit from the Berlingot occurs later and later. As a consequence the maximal obliquity increases more and more. Furthermore, from panel {\bf b} to {\bf c}, we see a modification of the topology of the intersection ${\cal S}\cap \partial {\cal B}$.  If we imagine ${\cal B}$ as an ocean and ${\cal S}$ as a continent, panel {\bf c} is a strait which gets larger as the companion semiminor axis decreases (panel {\bf d}). Once the strait is open, trajectories passing through it reach the retrograde side of the Berlingot and produce high spin-orbit misalignment. 
\figMaxTilta
\figMaxTiltb
The intersections between the two surfaces for each distance of the perturber have been computed with Maple. Then, for each point $(x, y, z)$ of the intersections, the initial condition $I'_0=\cos^{-1}z_0$ are computed from (\ref{eq.z0}), and the spin-orbit angle from $\theta_x = \cos^{-1}x$. The results are displayed in figure \ref{fig.maxtilta} for $0\leq\theta_z\leq 90^\circ$ and $0\leq I'_0\leq 90^\circ$. In Fig.~\ref{fig.maxtilta}.{\bf a}, the dashed curves represent the intersections ${\cal S}\cap \partial {\cal B}$ associated to the different values of the semiminor axis $b'$, and the shaded areas show the regions ${\cal S}\cap {\cal B}$ where the hyperboloid is inside the Berlingot. The darkest shaded areas, associated to the largest distances of the perturber, are disconnected. In these conditions, systems starting with zero spin-orbit angle must stay in the lowest regions and cannot reach misalignment larger than about 50$^\circ$.  On the contrary, when the strait is open, shaded areas become connected and at the vertical of the strait, systems reach high obliquities of the order of 100$^\circ$. Nevertheless, in the $(I'_0, \theta_x)$ plane, evolutions are vertical, thus at both sides of the strait systems conserve relatively low obliquities. Figure \ref{fig.maxtilta}.{\bf b} shows the maximal reachable spin-orbit misalignments deduced from Fig.~\ref{fig.maxtilta}.{\bf a}. Interestingly, we see that when the companion's semiminor axis is of the order of $170$ au (lightest gray area), even an inclination as small as 15 degrees is enough to generate a spin-orbit misalignment of about 100 degrees. 
It is worth noting that, in the limit where most of the angular momentum is in the orbit of the companion, the hyperboloid ${\cal S}$ and the Berlingot ${\cal B}$ are invariant by the transformation $(x,y,z)\rightarrow(x,-y,-z)$ according to (\ref{eq.approxS}) and (\ref{eq.Vxyz}). The spin-orbit angle $\theta_x$ is thus the same whether the orbit of the outer companion is prograde or retrograde.  Hence, Fig.~\ref{fig.maxtilta} stays unchanged if the abscissa $I'_0$ is replaced by $180^\circ-I'_0$.
Once the intersections between the Berlingot and the hyperboloid are known, the nutation periods $P_{\rm nut}$ associated to the oscillation of $\theta_x$ are obtained numerically from Eq.~(\ref{eq.Pn}). These periods are plotted in Fig.~\ref{fig.maxtilta}.{\bf c-f}. When the perturber is close ($b'=170,180$ au, Fig.~\ref{fig.maxtilta}.{\bf c,d}), $P_{\rm nut}$ shows spikes at the exact position of the borders of the strait.  As the perturber semiminor axis increases (Fig.~\ref{fig.maxtilta}.{\bf e,f}), the strait disappears as well as the spikes. The spikes are actually associated to trajectories along separatrices where the period is infinite. Indeed, these trajectories are tangent to the intersection ${\cal S} \cap \partial{\cal B}$ and thus, they are tangent to the surface of the Berlingot $\partial{\cal B}$. Moreover, from Fig.~\ref{fig.SB}.{\bf d}, it is manifest that the tangency is `from the inside', as in Fig.~\ref{fig.equil}.{\bf b}, so these trajectories actually pass through an hyperbolic fixed point. Besides those unstable fixed points, two stable fixed points exist in all panels of Fig.~\ref{fig.maxtilta}. They are at $I'_0=0$ and 90$^\circ$. In Fig.~\ref{fig.SB}, these initial conditions are located at the extremity $(1,1,1)$, and at the middle $(1,0,0)$, of the diagonal ${\cal D}_x$, respectively. In contrast to the previous fixed points, these ones belong to orbits which are tangent to the surface of the Berlingot `from the outside', as in Fig.~\ref{fig.equil}.{\bf a}. They are thus stable, and indeed, the amplitude of nutation at these two points (Fig.~\ref{fig.maxtilta}.{\bf b}) are nil.
To check the analytical results, we performed numerical integrations with $a'=170$ au and $e'=0$ starting at $1^\circ \leq I'_0 \leq 89^\circ$ by step of $1^\circ$. The results are shown in Fig.~\ref{fig.maxtilta}.{\bf b,c}.  The strait is actually present in the numerical simulations, although its width derived analytically was underestimated by about 10 degrees. The nutation period obtained numerically has the expected shape with two spikes at the border of the strait except that a small discrepancy exists inside the strait where the analytical method slightly overestimates the period. Nevertheless, the qualitative behavior is well recovered, and the quantitative differences must be due to the weak coupling between the inner planet and the outer ones as observed in Fig.~\ref{fig.Kaibc}.
The same exercise has been carried out on the system HD~20794, with $m'=1M_{\rm J}$ and $b'\in\{70, 60, 58.6, 55, 50\}$ au. Results are plotted in Fig.~\ref{fig.maxtiltb}. The two systems present a few qualitative similarities: 1) for large semiminor axis of the perturber, ${\cal S}\cap {\cal B}$ is disconnected, the inner system is quite decoupled from the outer companion and the stellar obliquity $\theta_x$ remains low whatever is the inclination of the perturber; 2) at small semiminor axis, a strait forms, spikes appear in the nutation period and some initial conditions lead to very high spin-orbit angle amplitudes. Nevertheless, behaviors are quantitatively different. In the case of the system 55~Cnc, the strait opens at low inclination $I'_0\sim 30^\circ$ and high values of $I'_0$ do not produce any strong spin-orbit misalignments, while for HD~20794, the strait is created at larger inclination $I'_0\sim 60^\circ$ and a high initial inclination is required to produce high stellar obliquity.  Moreover, in contrary to 55~Cancri, the maximal obliquity $\theta_x$ of HD~20794 never exceeds twice the initial inclination $I'_0$ (dash-dotted line in Fig.\ref{fig.maxtiltb}.{\bf b}). The origin of these differences will be discussed in the following. For this system, numerical simulations done at $b'=58.6$ au (Fig.~\ref{fig.maxtiltb}.{\bf b,e}) are in perfect agreement with the analytical results because the planets are well coupled.
\subsection{Spanning the parameter space}
The relative evolution described by the analytical model (section~\ref{sec.integrable}) depends on four parameters: the frequencies $\nu_1$, $\nu_2$, $\nu_3$, $\nu_4$ (see Eq.~(\ref{eq.evolxyz})). If we discard the temporal evolution, we can normalize all these frequencies by $\nu_4$, for example, and reduce the dimension of the parameter space to three. This dimension is still too large to be spanned entirely. Moreover, for each set of frequencies, we wish to analyze the curve representing the maximal obliquity of the star as a function of the initial inclination which adds one more dimension.
To solve this issue, we limit the study to asymptotic cases described as follows. Let $(\nu_a, \nu_b, \nu_c, \nu_d)$ be a permutation of $(\nu_1, \nu_2, \nu_3, \nu_4)$. We distinguish 8 classes A, B, C, D, E, F, G, H of parameters defined by
\begin{description}
\item[class A]: $ \nu_a \ll   \nu_b  \ll   \nu_c  \ll  \nu_d $\ ,
\item[class B]: $(\nu_a \sim  \nu_b) \ll   \nu_c  \ll  \nu_d $\ ,
\item[class C]: $ \nu_a \ll  (\nu_b  \sim  \nu_c) \ll  \nu_d $\ ,
\item[class D]: $ \nu_a \ll   \nu_b  \ll  (\nu_c  \sim \nu_d)$\ ,
\item[class E]: $(\nu_a \sim  \nu_b  \sim  \nu_c) \ll  \nu_d $\ ,
\item[class F]: $(\nu_a \sim  \nu_b) \ll  (\nu_c  \sim \nu_d)$\ ,
\item[class G]: $ \nu_a \ll  (\nu_b  \sim  \nu_c  \sim \nu_d)$\ ,
\item[class H]: $ \nu_a \sim  \nu_b  \sim  \nu_c  \sim \nu_d $\ .
\end{description}
The number of distinct permutations in class A is $n({\rm A})=4!=24$.  In class B, there are half as many distinct permutations because switching $\nu_a$ and $\nu_b$ does not change the order of the frequencies. Thus, $n(\rm B)=12$. Similarly, $n({\rm C}) = n({\rm D}) = 12$. Using simple enumerative combinatorics, one gets $n({\rm E})=4$, $n({\rm F})=6$, $n({\rm G})=4$, and $n({\rm H})=1$. Adding these cases all together only leads to 75 different configurations. 
We analyze them all. For each pair of frequencies satisfying $\nu_a\ll \nu_b$, we take $\nu_b = 10 \nu_a$, and the cases $\nu_a\sim\nu_b$ are replaced by exact equalities $\nu_a = \nu_b$. As a result, we identify four different regimes.
\subsubsection{Cassini regime}
We call systems in the Cassini regime those satisfying $(\nu_2,\nu_4)\ll(\nu_1\sim\nu_3)$. This hierarchy of frequencies implies that 1) the planet system is not affected by the orientation of the central star ($\nu_2$ is small); 2) the plane of the outer orbit is fixed ($\nu_4$ is small); and 3) the precession frequency of the star matches that of the planet system ($\nu_1\sim \nu_3$). These hypothesis are those leading to the well-known Cassini problem \citep[e.g.,][]{Colombo_AJ_1966, Peale_AJ_1969, Henrard_Murigande_CeMDA_1987, Ward_Hamilton_AJ_2004}. Among the 75 different configurations, three of them fulfill the condition $(\nu_2,\nu_4)\ll(\nu_1\sim\nu_3)$, with $\nu_2\ll \nu_4$, $\nu_2\sim \nu_4$, or $\nu_4\ll\nu_2$. They all present similar nutation amplitudes $\theta_{x,\rm max}$ as a function of the inclination $I'_0$ (Fig.~\ref{fig.regimes}.{\bf a}). Interestingly, the maximal stellar obliquity relative to the planet plane exceeds twice the initial inclination $I'_0$ between the planet plane and the outer orbit.
\figRegimes
As an example, the system 55~Cnc where the perturber is put on a rather close-in orbit $b'\lesssim 170$ au belongs to the Cassini regime with $\nu_4\ll \nu_2 \ll (\nu_1\sim \nu_3)$. The dynamics of this type of system is usually analyzed in a frame rotating at the precession frequency (as in Figs.~\ref{fig.Kaibb}.{\bf d}, \ref{fig.Kaibc}.{\bf d}, \ref{fig.HDa}, and \ref{fig.HDb}), by plotting the projected trajectories of the spin-axis on the invariant plane for different initial obliquities.  Numerical integrations of the 55~Cnc system with $b'=170$ au and $I'_0=15^\circ$ are displayed in Fig.~\ref{fig.cassini}. To make the comparison easier, the axes are the same as in \citet{Ward_Hamilton_AJ_2004}. Trajectories in Fig.~\ref{fig.cassini} present small amplitude oscillations which are due to the individual motion of the orbit of each planet. Nevertheless, Cassini states 1, 2, and 4  are easily identifiable, and the large increase of the obliquity observed in Figs.~\ref{fig.SB} and \ref{fig.regimes}.{\bf a} is due to the wide oscillation of $\vec s$ around the Cassini state 2 along the trajectory passing through the point labeled $\vec w$ in Fig.~\ref{fig.cassini} (the initial condition $\theta_x(0)=0$ implies $\vec s(0)=\vec w(0)$).
\figCassini
\subsubsection{Pure orbital regime}
When the frequencies associated to the spin-orbit coupling $\nu_1$ and $\nu_2$ are much smaller than those of the planet-companion interaction $\nu_3$ and $\nu_4$, the evolution of the system becomes purely orbital. The stellar orientation remains fixed and does not perturbed the motion of the orbital planes. According to the distribution of angular momentum between the planets and the companion, this regime leads to different spin-axis evolutions displayed in Fig.~\ref{fig.regimes}.{\bf b,c,k}. 
If the companion contains most of the angular momentum, its orbital plane is almost invariant and the planet system precesses around it at constant inclination. In this case, $(\nu_1, \nu_2, \nu_4)\ll\nu_3$ and the spin-orbit angle oscillates  between 0 and ${\rm min}(I'_0, 180^\circ-I'_0)$ (Fig.~\ref{fig.regimes}.{\bf b}). The period associated to this motion is simply $P_{\rm nut}=2\pi/\nu_3$.  In this regime, we can cite in particular Kepler-56, the only multiplanet system with an observed spin-orbit misalignment \citep{Huber_etal_Science_2013}. Also, the five planets of the 55~Cancri system, with the outer one playing the role of the perturber and the others coupled together, lies in this regime. In that case, the inclination of the innermost planet with respect to the plane of the sky is known by transit to be $82.5\pm1.4^\circ$ \citep{Gillon_etal_AA_2012} while the outermost is measured at $53\pm7^\circ$ by astrometry \citep{McArthur_etal_ApJ_2004}.  According to these observations, the orbits should be mutually inclined by at least 20$^\circ$ which would give rise to a periodic misalignment of $\gtrsim 40^\circ$.
When the angular momentum is equally distributed between the planets and the companion, $\nu_3\sim\nu_4$ and both orbital planes precess around the total orbital angular momentum $\vec G+\vec G'$. As a consequence, if the initial obliquity $\theta_x(0)$ is nil, the two angles $\theta_x$ and $\theta_y$ oscillate in antiphase between 0 and $I'_0$ (Fig.\ref{fig.regimes}.{\bf c}). The corresponding nutation period is $P_{\rm nut} = \pi/(\nu_3 \cos I'_0 \cos(I'_0/2))$.
In the last case, when planets have significantly more angular momentum than the companion, the orbital frequencies satisfy $\nu_3\ll\nu_4$.  As a result, the planets plane does not move while the outer body precesses around it at the frequency $\nu_4$. Once the latter motion is averaged out, the remaining interaction in the system is between the stellar equatorial bulge and the planets, but the angular momenta are aligned, so this configuration does not produce any spin-orbit misalignment (Fig.~\ref{fig.regimes}.{\bf k}).  Such would be the case if a hot Jupiter were perturbed by an Earth analogue.
More generally, within the pure orbital regime and considering all orbital angular momentum distributions, the amplitude of the spin-orbit angle $\theta_x$ varies from 0 to $2I'_0$ as $G'/G$ increases from zero to infinity, the mutual inclination $\theta_z$ remains constant equal to $I'_0$, and the associated nutation period is
\be
P_{\rm nut} = \frac{2\pi G' G}{\gamma\cos\theta_z \sqrt{G'^2+G^2+2G'
G\cos\theta_z}}\ .
\ee
\subsubsection{Laplace regime}
We now consider the case where the frequency $\nu_2$ is one of the largest frequencies in the system, but $\nu_1$ remains much smaller.  This occurs only if the star has more angular momentum than the planets ($L\gg G$). With these parameters, the stellar orientation does not evolve and the motion of the planet plane is dictated by the inclination of the Laplace surface with respect to the equator of the star. The Laplace surface, commonly used in satellite problems, is such that inclinations measured relative to it remain roughly constant (\citetalias{Boue_Laskar_Icarus_2006}; \citealt{Tremaine_etal_AJ_2009}).  For close-in planets with $\nu_2\gg \nu_3$, the Laplace surface is aligned with the stellar equator and the spin-orbit angle remains constant (Fig.~\ref{fig.regimes}.{\bf k}). In systems with longer period planets satisfying $\nu_3\gg(\nu_2,\nu_4)$, the Laplace surface is that of the companion's orbit.  The latter configuration falls into the above mentioned pure orbital regime characterized by $\theta_{x,\rm max} = I'_0$. In between the two extremes ($\nu_2\sim\nu_3$, Fig.~\ref{fig.regimes}.{\bf d,e}), the Laplace surface is situated half way between the stellar equator and the outer orbit.  In particular, when $(\nu_1,\nu_4)\ll(\nu_2\sim\nu_3)$ (Fig.~\ref{fig.regimes}.{\bf d}), $\theta_x$ and $\theta_z$ oscillate in antiphase between 0 and ${\rm min}(I'_0, 180^\circ-I'_0)$. The evolution is more complicated when all the three frequencies $\nu_2$, $\nu_3$, and $\nu_4$ are similar (Fig.~\ref{fig.regimes}.{\bf e}).
\subsubsection{Hybrid regime}
The hybrid regime gather systems in which the stellar precession frequency $\nu_1$ is larger than or of the same order as the dominant orbital one(s).  Thus, both the equator of the star and the orbits have a nutation motion. The Cassini regime $(\nu_2,\nu_4)\ll(\nu_1\sim\nu_3)$ fulfills this criterion, but given its importance for spin-orbit misalignment, it has been studied separately. Other configurations are $\nu_2\ll\nu_3\ll(\nu_1\sim\nu_4)$ (Fig.~\ref{fig.regimes}.{\bf f}), $(\nu_2\sim\nu_3)\ll(\nu_1\sim\nu_4)$ (Fig.~\ref{fig.regimes}.{\bf g}), $\nu_2\ll(\nu_1\sim\nu_3\sim\nu_4)$ (Fig.~\ref{fig.regimes}.{\bf h}), $\nu_4\ll(\nu_1\sim\nu_2\sim\nu_3)$ (Fig.~\ref{fig.regimes}.{\bf i}), and $\nu_1\sim\nu_2\sim\nu_3\sim\nu_4$ (Fig.~\ref{fig.regimes}.{\bf j}).  These cases lead to intermediate spin-orbit angle amplitudes. All the others do not produce any significant misalignment (Fig.~\ref{fig.regimes}.{\bf k}).  For example, using the parameters of \citet{Kaib_etal_ApJ_2011}, we obtained $\nu_1\gg(\nu_2,\nu_3,\nu_4)$ which is a particular case of the hybrid regime that does not produce any increase of the stellar obliquity.
\section{Conclusion}
The dynamics of compact multiplanet systems perturbed by an outer companion reveals itself very rich. These systems can be represented by three unit vectors $\vec s$, $\vec w$, and $\vec w'$ along the angular momentum of the star, of the planet system, and of the companion's orbit, respectively. Their dynamics is mainly the combination of a uniform rotation of the three vectors around the total angular momentum and of a periodic motion in the rotating frame. These two evolutions are called precession and nutation in reference to the Earth-Moon problem perturbed by the Sun. The relative motion described in the frame $(x,y,z)$, with $x=\vec s\cdot \vec w$, $y=\vec s\cdot \vec w'$, and $z=\vec w\cdot \vec w'$, is integrable whatever are the four precession constants $\nu_1$, $\nu_2$, $\nu_3$, $\nu_4$ (see Eqs.~(\ref{eq.nus})) involved in the problem. The solutions are derived geometrically as in \citetalias{Boue_Laskar_Icarus_2006} and \citetalias{Boue_Laskar_Icarus_2009}. 
In this work, we went even further in the geometrical approach. In particular, gathering all evolutions starting with a spin-orbit alignment and an arbitrary inclination of the companion, nutation amplitudes (and in particular the amplitude of the spin-orbit angle) are obtained all together without numerical simulations from the intersection of a Berlingot shaped surface defined in \citetalias{Boue_Laskar_Icarus_2006} with an hyperboloid of one sheet equal to the union of all trajectories in the $(x,y,z)$ frame. The parameter space of the problem controlled by the precession constants is three-dimensional. It is thus not possible to explore the whole dynamics. Nevertheless, using simple enumerative combinatorics, we identified 75 different asymptotic configurations.  We computed the nutation amplitudes on all these cases and distinguished four different regimes: 
\begin{enumerate}
\item the Cassini regime where the stellar spin evolution is well described by the Cassini approximation and the obliquity can exceed twice the initial inclination of the perturber. This occurs when the star has much less angular momentum than the planets and the companion; 
\item the pure orbital regime where the stellar orientation remains fixed and does not perturb the orbital evolution. In that case, due to the motion of the planet system, the stellar nutation amplitude varies between zero and twice the initial inclination of the perturber as the orbital angular momentum ratio of the latter over that of the planet system increases from zero to infinity;
\item the Laplace regime where the stellar spin-axis is fixed and planets precess around a Laplace plane between the equator of the star and the orbit of the companion. In this regime, the amplitude of the spin-orbit angle varies between zero and the initial inclination of the perturber. The maximum is reached when the equatorial bulge and the outer body produce similar torques on the planets; 
\item the hybrid regime where the precession rate of the stellar equator is larger than or similar to those of the orbit planes. Systems in the Cassini regime are excluded since they have already been studied.  The hybrid regime is characterized by intermediate nutation amplitudes ranging from zero to almost twice the initial inclination of the outer body.
\end{enumerate}
This formalism applied to the 55 Cancri system showed that the central star is linked to the planets unless the outer companion is on a very high eccentric orbit ($e'\gtrsim 0.95$). Indeed, most of the angular momentum is in the planets and the evolution of the stellar spin-axis is dictated by the precession rate of the planet system around the binary orbit plane. Below the eccentricity limit the perturbation by the outer body is weak enough that the stellar spin-axis follows the planet motion. But close to the threshold, the system is in the Cassini regime characterized by the resonance between the stellar and the planetary precession frequencies. In that case, the stellar obliquity exceeds twice the inclination of the companion.
The geometrical analysis has been made possible thanks to the vectorial formalism detailed in BF14.  Thus we have shown spin-orbit mechanics, in which the orbital mechanics are far from trivial, as a particular application of that approach. 
Taking both papers together, we first provided the formalism to analyze dynamically rigid planet systems interacting with the spin-axis of their parent star and with an outer companion; and then we gave a general view of all possible evolutions. The next step is to develop and analyze different scenarios which may produce spin-orbit misalignment according to the results of this study.  Particularly, we expect that evolution in the spin rate of the star or orbital migration may excite spin-orbit resonances which leave the star tilted.  We also hope the present formalism may reveal deeper understanding of trends in spin-orbit measurements, or at least interesting application to individual systems with such measurements.
\acknowledgments
This work benefited a lot from GB's PhD thesis done with Jacques Laskar during which the three vector problem has been solved geometrically for the first time. GB is thus particularly grateful to Jacques Laskar. GB also thanks Philippe Robutel and Alain Chenciner for many helpful discussions about the vectorial approach and Cayley's nodal cubic surface.
\bibliographystyle{apj}
\bibliography{bf}
\end{document}